\documentclass[journal]{IEEEtran}
\usepackage[T1]{fontenc}

\def\argmax{\mathop{\rm \arg\!\max}}

\usepackage{graphicx}
\usepackage[noadjust]{cite}
\usepackage{mcite}
\usepackage{amsfonts,helvet}
\usepackage{fancyhdr}
\usepackage{threeparttable}
\usepackage{epsf,epsfig}
\usepackage{amsthm}
\usepackage{amsmath}
\usepackage{siunitx}
\usepackage{amssymb}
\usepackage{dsfont}
\usepackage{subfigure}
\usepackage{color}
\usepackage[linesnumbered,ruled]{algorithm2e}
\usepackage{algpseudocode}
\usepackage{algcompatible}
\usepackage{enumerate}
\usepackage{cancel}
\usepackage{bbm}
\usepackage{dsfont}
\usepackage[subnum]{cases}
\usepackage{adjustbox}
\usepackage[linesnumbered,ruled]{algorithm2e}
\usepackage{multicol}
\usepackage[english]{babel}
\usepackage{array}
\usepackage{lipsum}
\usepackage{textcomp}

\def\BibTeX{{\rm B\kern-.05em{\sc i\kern-.025em b}\kern-.08em
    T\kern-.1667em\lower.7ex\hbox{E}\kern-.125emX}}

\newtheorem{remark}{Remark}

\newmuskip\

\makeatletter
\def\blfootnote{\xdef\@thefnmark{}\@footnotetext}
\makeatother

\def\ba{{\bf a}}
\def\bbb{{\bf b}}
\def\bc{{\bf c}}
\def\bd{{\bf d}}

\def\bh{{\bf h}}

\def\bm{{\bf m}}

\def\br{{\bf r}}
\def\bs{{\bf s}}

\def\bu{{\bf u}}

\def\bw{{\bf w}}

\def\by{{\bf y}}
\def\bz{{\bf z}}

\def\bA{{\bf A}}

\def\bH{{\bf H}}
\def\bI{{\bf I}}

\def\bP{{\bf P}}

\def\bW{{\bf W}}

\def\cB{\mbox{$\mathcal{B}$}}
\def\cC{\mbox{$\mathcal{C}$}}

\def\cI{\mbox{$\mathcal{I}$}}

\def\cK{\mbox{$\mathcal{K}$}}

\def\cN{\mbox{$\mathcal{N}$}}

\def\cQ{\mbox{$\mathcal{Q}$}}

\def\cS{\mbox{$\mathcal{S}$}}

\def\bbC{\mbox{$\mathbb{C}$}}

\def\bbE{\mbox{$\mathbb{E}$}}

\def\bbP{\mbox{$\mathbb{P}$}}
\def\bbQ{\mbox{$\mathbb{Q}$}}

\def\bb1{\mathbbm{1}}

\def\bSigma{\boldsymbol{\Sigma}}

\begin{document}
\title{
Learning-Based One-Bit Maximum Likelihood Detection for Massive MIMO Systems: Dithering-Aided Adaptive Approach
}
\author{
Yunseong Cho, {\it Member, IEEE,} Jinseok Choi, {\it Member, IEEE,} 
\\ and Brian L. Evans, {\it Fellow, IEEE}   
\thanks{
Copyright (c) 2015 IEEE. Personal use of this material is permitted. However, permission to use this material for any other purposes must be obtained from the IEEE by sending a request to pubs-permissions@ieee.org.
}
\thanks{
Y. Cho is affliated with Samsung Research America, Plano, TX 75024, USA. (e-mail: 
y.cho1@samsung.com).

J. Choi is with Korea Advanced Institute of Science and Technology (KAIST), Daejeon, 34141, Republic of Korea
(e-mail:jinseok@kaist.ac.kr).

B. L. Evans is with the 6G@UT Research Center, Wireless Networking and Communication Group (WNCG), Department of Electrical and Computer Engineering, The University of Texas at Austin, Austin, TX 78701, USA. (e-mail: bevans@ece.utexas.edu)
}
\thanks{
This work was supported in part by NVIDIA and AT\&T Labs, affiliates of the 6G@UT Research Center at UT Austin, by BK21 FOUR (Connected AI Education \& Research Program for Industry and Society Innovation, KAIST EE, No. 4120200113769), by the National Research Foundation of Korea (NRF) grant funded by the Korea government (MSIT) (No. RS-2023-00219443), and by the MSIT (Ministry of Science and ICT), Korea, under the ITRC (Information Technology Research Center) support program (IITP-2023-RS-2023-00259991) supervised by the IITP (Institute for Information \& Communications Technology Planning \& Evaluation).
}
}




\maketitle

\begin{abstract}
In this paper, we propose a learning-based detection framework for uplink massive multiple-input and multiple-output (MIMO) systems with one-bit analog-to-digital converters. 
The learning-based detection only requires counting the occurrences of the quantized outputs of -1 and +1 for estimating a likelihood probability at each antenna.
Accordingly, the key advantage of this approach is to perform maximum likelihood detection without explicit channel estimation which has been one of the primary challenges of one-bit quantized systems.
However, due to the quasi-deterministic reception in the high signal-to-noise ratio (SNR) regime, one-bit observations in the high SNR regime are biased to either $+1$ or $-1$, and thus, the learning requires excessive training to estimate the small likelihood probabilities.
To address this drawback, we propose a dither-and-learning technique to estimate likelihood functions from dithered signals.
{\color{black} First, we add a dithering signal to artificially decrease the SNR and then infer the likelihood function from the quantized dithered signals by using an SNR estimate derived from a deep neural network-based estimator which is trained offline.}
We extend our technique by developing an adaptive dither-and-learning method that updates the dithering power according to the patterns observed in the quantized dithered signals. 
The proposed framework is also applied to  channel-coded MIMO systems by computing a bit-wise and user-wise log-likelihood ratio from the refined likelihood probabilities.
Simulation results validate the  performance of the proposed methods in both uncoded and coded systems.
\end{abstract}

\blfootnote{This work was presented in part at the {\em IEEE Global Communications Conference (IEEE GLOBECOM)}, Waikoloa, Hawaii, 2019 \cite{choi2019robust}.}

\section{Introduction}
\label{sec:intro}

Massive MIMO systems for sub-6 GHz wireless communications \cite{ngo2013energy, larsson2014massive} and millimeter wave (mmWave) communications \cite{pi2011introduction, andrews2014will, heath2018foundations} have been considered as one of the emerging technologies for future communications because of the remarkable improvements in terms of spectral efficiency and capacity gains \cite{marzetta2010noncooperative}.
As wireless communication systems continue to grow in popularity and become increasingly important, there is a growing need to investigate communication systems that are not only reliable and high-performing, but also energy-efficient for various future wireless applications such as vehicle-to-everything, internet-of-things, extended reality, and smart grid~\cite{saad2019vision,de2021survey}.
The small wavelength of mmWave signals and the reduced antenna spacing in mmWave systems enable the installation of more antennas per unit area. Each of these antennas is connected to a dedicated radio frequency (RF) chain equipped with a pair of high-precision data converters which can unlock enhanced spatial coverage and improved signal processing capabilities.

However, the use of a large number of high-resolution analog-to-digital converters (ADCs) at receivers results in prohibitively huge power consumption, which becomes the main bottleneck in the practical deployment because a high-resolution ADC is particularly power-hungry as the power consumption of an ADC tends to scale up exponentially with the number of quantization bits.
To overcome the circuit power issue, deploying low-precision ADCs has been considered as a promising low-power solution over the past years \cite{wen2016bayes,studer2016quantized,choi2017resolution,choi2018antenna,choi2021quantized}. 
As an extreme case of the low-resolution data converters, the use of one-bit data converters has emerged and become particularly attractive due to the ability to dramatically enhance power efficiency, lower hardware cost, and simplify analog processing of receivers \cite{mezghani2007ultra, mo2015capacity,park2021construction,wang2015multiuser, choi2015quantized, choi2016near,cho2019one,mollen2017uplink}.
Because of the strong nonlinearity, data detection and channel estimation with one-bit data converters are known to be more challenging; however, the use of massive antenna arrays can alleviate the performance loss 
\cite{shao2017iterative, li2017downlink}.
Nevertheless, when conventional signal processing algorithms are applied directly to low-resolution systems, significant performance losses can occur due to the severe nonlinear distortions that low-resolution ADCs cause.

State-of-the-art one-bit  detection, beamforming, and channel estimation techniques have been developed in the recent decades~\cite{park2021construction,wang2015multiuser, choi2015quantized, choi2016near,cho2019one,liu2021efficient,jacobsson2017throughput}.
Low-complexity symbol-level beamforming methods for one-bit quantized systems were developed for quadrature-amplitude-modulation (QAM) constellations \cite{park2021construction}.
Taking into account the heavily quantized signals and antenna correlations,  an iterative multiuser detection powered by a message-passing de-quantization algorithm was devised in  \cite{wang2015multiuser}.
In \cite{choi2015quantized}, a high-complexity one-bit ML detection and low-complxity zero-forcing (ZF)-type detection methods were developed.
In terms of MIMO detectors, by converting the ML estimation problem in \cite{choi2015quantized} to convex optimization, the optimal maximum-likelihood (ML) detector was introduced and the near-ML detector was also proposed by transforming the ML detection problem into a tractable convex optimization problem \cite{choi2016near}.
Successive-interference-cancellation one-bit receivers that can be applied to modern channel coding techniques was presented in \cite{cho2019one}.
Machine learning techniques were also employed for one-bit detection \cite{nguyen2021svm,balevi2019one,nguyen2021linear}.
It was shown in \cite{nguyen2021svm} that support vector machines can be used for efficient channel estimation and data detection with one-bit quantized observations.
In \cite{balevi2019one}, the conventional orthogonal frequency division multiplexing precoder and decoder are replaced with artificial neural networks to enable unsupervised autoencoder-based detection.
The authors in \cite{nguyen2021linear} combined a linear estimator based on the Bussgang decomposition and a model-based deep neural network (DNN) approach to make data detection with one-bit ADCs adaptive to the current channel.
Although the aforementioned state-of-the-art one-bit detectors provide high detection performance, the detection methods require the estimation of channel state information (CSI) which is one of the key challenges in one-bit quantized communication systems.

Accordingly, numerous one-bit ADC channel estimation methods have been developed such as least-squares (LS), ML, and Bussgang decomposition-based methods \cite{choi2016near,li2017channel}.
Combined with antenna-wise non-zero thresholding for one-bit data quantizers, the majorization-minimization-based ML channel estimator was proposed in \cite{liu2021efficient}.
In \cite{jacobsson2017throughput}, it was shown that Bussgang decomposition-based channel estimator can provide reliable performance for high-order constellations in one-bit ADC systems.
The authors in \cite{zhang2020deep} utilized supervised deep learning in developing a mapping from the one-bit quantized measurements to the wireless channels.
The authors in \cite{rao2019channel} derived the lower bounds on the performance of the channel estimation in one-bit MIMO systems
considering various mmWave channel models.
These recent advancements in channel estimation schemes for one-bit quantized signals, however, still suffer degradation in estimation accuracy compared with high-precision ADC systems.

{\color{black} Dithering has found application in one-bit ADC systems for various purposes. 
In \cite{mezghani2020massive}, dithering served the purpose of mitigating correlations within spatial quantization errors for sub-wavelength spatial sampling. 
Essentially, the application of dithering before quantization was intended to decorrelate distortion errors—a crucial aspect for achieving ideal performance through linear processing. 
In \cite{liang2016mixed}, it was demonstrated that a Gaussian-type dither can enhance the effective bit-width of one-bit ADCs, thereby aiding in the reduction of estimation bias for channel estimation.
Other one-bit ADC works incorporating dithering signals also acknowledge the utility of dithered quantizers \cite{papadopoulos2001sequential}. 
For instance, in the case of a Gaussian prior on channel coefficients, the use of the linear minimum mean squared error estimate of the channel as a dither signal is demonstrated to be effective in practice \cite{dabeer2010channel}.
}

As another research direction, learning-based data detection techniques have recently been investigated to remove or minimize the requirement for explicit channel estimation in one-bit ADC MIMO systems \cite{jeon2018one,jeon2018supervised,jeon2018reinforcement,park2023learning}.
The authors in \cite{jeon2018one} applied sphere decoding to the one-bit quantized systems and showed that the detection complexity can be reduced while achieving near-optimal performance.
Viewing the one-bit ADC systems as a classification problem, various supervised-learning-based data detection techniques were provided by estimating effective channels and learning the non-linear system response \cite{jeon2018supervised}.
In \cite{jeon2018reinforcement}, however, a channel estimation was done to initialize likelihood functions for ML detection, and a learning-based likelihood function was used for post-update of the likelihood functions.
In contrast, the authors in \cite{park2023learning} used an estimated channel to generate noisy training pilots and developed an expectation-maximization algorithm that facilitates the likelihood probability learning process. 
Unlike previous learning-based approaches that focused on developing detection mechanisms based on estimated channels, we focus on applying one-bit ML detection and learning likelihood functions without channel estimation.
{\color{black}Then we propose a novel dithering-based learning to overcome the problem of the learning process with the limited amount of training.
We remark that  the primary goals associated with the dithering signal in previous works \cite{mezghani2020massive,liang2016mixed,papadopoulos2001sequential,dabeer2010channel} differ from the specific objective of our work.}


\subsection{Contributions}
In this work, we explore a learning-based ML detection approach that replaces a one-bit channel estimation stage with a counting-based learning process for an uplink multiuser MIMO systems with one-bit ADCs.
The contributions of this work are summarized as
follows:
\begin{itemize}
\item We propose a dither-and-learning technique to infer the likelihood functions from dithered signals. Such an approach significantly reduces the number of zero-valued likelihood functions experienced by naive learning-based one-bit detection. 
After the dithering process, we obtain a preferable statistical pattern in the one-bit quantized output sequences with moderate sign changes thanks to the reduced SNR.
Then a denoising phase retrieves the actual likelihood functions without the impact of the dithering noise.
The proposed method allows estimating the likelihood functions with a reasonable training length by drawing meaningful sign patterns in the quantized output sequence.
\item To further improve  learning accuracy, we develop an adaptive dither-and-learning technique for adjusting each antenna element's dithering power according the patterns observed
in the quantized dithered signals.
Since the performance of the proposed dithering-based learning algorithm is affected by the dithering power, the proposed feedback-based adaptive algorithm effectively adjusts the dithering noise power depending on the pattern of the one-bit quantized outputs.
{\color{black} A DNN-based SNR estimation method is also developed to facilitate the denoising phase of the dithering-based learning in the practical systems.}

\item To further apply the proposed learning-based scheme to modern communication frameworks rather than being limited to hard-output detection, we compute the log-likelihood ratio (LLR), i.e., soft output, which is then fed into a channel-decoder. 
Noting that the LLR needs to be defined with respect to an individual binary bit of each user, we separate the index set of all possible symbol vectors into two disjoint subgroups and compare the sum of the likelihood probabilities over the two subgroups.
\item Simulation results validate that, in contrast to the conventional learning-based one-bit ML detectors and other channel estimation-based one-bit data detectors, the proposed detectors can achieve comparable performance to the optimal one-bit ML detection that operates with perfect CSI and exhibit more reliable detection performance in both uncoded and coded simulation cases.
\end{itemize}

This paper is organized as follows. 
In Section \ref{sec:system}, we introduce the uplink MIMO signal model and the optimal one-bit ML detection rule. 
Section \ref{sec:ML_learning} provides a counting-based one-bit ML detection strategy that does not require channel estimation. 
In Section \ref{subsec:DL}, we propose the learning-based ML detection, using dithering noise to relax the limitation of the counting-based approach.
Section \ref{subsec:ADL} explores the adaptation of the dithering noise variance
and Section \ref{subsec:SNRestimation} delivers a DNN-based SNR estimation needed for the de-noising stage.
We extend the proposed ML mechanism to channel-coded communication systems in Section \ref{sec:coded}.
In Section \ref{sec:simul}, the proposed detection methods are evaluated.
Section \ref{sec:con} concludes the paper.

{\it Notation}: $\bf{A}$ is a matrix and $\bf{a}$ is a column vector. 
$\mathbf{A}^T$ and $\ba^T$ denote the transpose operation of matrix and column vector, respectively. 
We denote $a_{i}$ as the $i$th element of $\bf a$. 
With mean $\mu$ and variance $\sigma^2$, we generate a real Gaussian distribution and a complex Gaussian distribution using $\mathcal{N}(\mu, \sigma^2)$ and $\mathcal{CN}(\mu, \sigma^2)$, respectively. 
${\rm {diag}} (\bf a)$ creates a diagonal matrix that has $a_i$'s as its diagonal entries. 
${\bf 1}_N$ and ${\bf 0}_N$ are a $N \times 1$ one vector and zero vector, respectively.
$\bI_{N}$ denotes the $N\times N$ identify matrix.
${\rm {Re}}\{\bA\}$ and ${\rm {Im}}\{\bA\}$ take the real and imaginary part of $\bA$, respectively.
$\bb1{\{A\}}$ denotes the indicator function which outputs 1 if $A$ is true, and 0 otherwise.
$\bbP[\cdot]$ and $\bbE[\cdot]$ are the probability and expectation operators, respectively.

\section{System Model}
\label{sec:system}

In this section, we describe the uplink MIMO system model and the optimal one-bit ML detection rule which is feasible in the case of perfect CSI. 

\subsection{Signal Model}
\label{sec:signal model}

We consider uplink multiuser MIMO communication systems where the base station (BS) equipped with $N_r$ receive antennas concurrently communicates with $N_u$ single-antenna users.
We suppose $N_r \gg N_u$ in the context of massive MIMO systems.
Each antenna element has its own dedicated RF chain as well as individual in-phase and quadrature one-bit ADCs.
The wireless channel follows a block fading model whose channel matrix is invariant for $N_c$ coherent time slots.
We then split the uplink transmission into a training phase with $N_t$ time slots and a data transmission phase with $N_d$ slots, i.e., $N_c=N_t+N_d$.

During the training phase, each user transmits up to $N_t$ pilot symbols.
We use $K$ to denote the number of possible pilot symbol combinations of  $N_u$ users.
{\color{black} When all users adopt an $M$-ary constellation, we have $K = M^{N_u}$, e.g., $K=2^{N_u}$ for binary phase shift keying at all $N_u$ users.
In the considered system, each pilot symbol combination is transmitted $N_{\rm tr}$ times, which implies that the condition of $N_t = KN_{\sf tr}$ is necessary} to capture the characteristics of all possible combinations.
{\color{black} 
Accordingly, to reduce the overall training overhead, it is desirable to reduce both $N_{\rm tr}$ and $K$.
In this paper, we focus on improving the learning-based one-bit ML detection performance by reducing the training repetition for each candidate vector, i.e. $N_{\rm tr}$, and  we leave the problem of reducing $K$ as a future work.}

The set of the constellation points of $M$-ary QAM scheme is represented by $\bbQ_M$, from which $\bar s_u[t]$ is generated where $\bar s_u[t]$ is the complex-valued QAM data symbol of the $u$th user at time $t$.
We assume that $\bar s_u[t]\in\bbQ_M$ has zero mean and unit variance, i.e.,  $\bbE[\bar s_u] =0$ and $\bbE[|\bar s_u[t]|^2] = 1$.
A symbol vector $\bar \bs[t] = \big[\bar s_1[t], \dots, \bar s_{N_u}[t]\big]^T \in \bbQ_M^{N_u}$, $t \in \{ 1,\dots, N_c\}$ denotes the collection of the transmitted signals from $N_u$ users at time $t$.
We consider each user to adopt $M$-ary QAM constellation and thus, the total number of possible symbol vectors $\bar \bs[t]$ becomes $K=M^{N_u}$ which is the cardinality of $\bbQ_M^{N_u}$.
Assuming that the transmitted symbols from users are concurrently received and jointly processed at the BS, the received analog complex baseband signal vector at time $t$ can be represented as 
\begin{align}
    \label{eq:r}
    \bar\br[t] = \sqrt{\rho}\bar\bH^T\bar\bs[t] + \bar\bz[t],
\end{align}
where $\bar\bH \in \bbC^{N_u\times N_r}$ is the complex-valued channel matrix between the BS  and $N_u$ users, whose $i$th column vector, i.e.,  $\bar\bh_i$, indicates the uplink channel vector defined for the propagation from all user to the $i$th antenna element of the BS.
The transmit power is denoted as $\rho$, and the additive white complex Gaussian noise vector ${\bar \bz}[t]$ follows ${\bar \bz}[t]\sim \cC\cN({\bf 0}_{N_r}, N_0\bI_{N_r})$ where $N_0$ is the AWGN noise variance.
Here, we define the SNR as 
\begin{align}
    \gamma = \rho/N_0.
\end{align}

Then, each real and imaginary component of the received signals in \eqref{eq:r} is quantized by one-bit ADCs which only reveal the sign of the signals, i.e., either $+1$ or $-1$.
The complex-valued quantized signal can be represented as
\begin{align}
    \bar\by[t] = \cQ({\rm {Re}}\{\bar \br[t]\}) + j\cQ({\rm {Im}}\{\bar \br[t]\})
\end{align}
where $\cQ(x)=(-1)^{\bb1\{x\leq0\}}\in\{-1,+1\}$ is an element-wise one-bit data quantizer which returns $+1$ if the input is positive, or $-1$ otherwise.
The received signal in the complex-vector expression $\bar \br[t]$ can be rewritten in a real-valued vector representation as
\begin{align}
    \br[t] &= \begin{bmatrix} {\rm {Re}}\{\bar\br[t]\} \vspace{0.3em} \\  {\rm {Im}}\{\bar\br[t]\}\end{bmatrix} \\
    \label{eq:rx signal}
    &= \sqrt{\rho}\bH^T\bs[t] + \bz[t]
\end{align}
where 
\begin{gather}
    \bH^T = \begin{bmatrix} {\rm {Re}}\{\bar\bH^T\} & -{\rm {Im}}\{\bar\bH^T\} \vspace{0.3em} \\ {\rm {Im}}\{\bar\bH^T\} & {\rm {Re}}\{\bar\bH^T\}\end{bmatrix},    \\
    \bs[t] = \begin{bmatrix} {\rm {Re}}\{\bar\bs[t]\} \vspace{0.3em}\\  {\rm {Im}}\{\bar\bs[t]\}\end{bmatrix}, \label{eq:real_x}\\
    \bz[t] = \begin{bmatrix} {\rm {Re}}\{\bar\bz[t]\} \vspace{0.3em}\\  {\rm {Im}}\{\bar\bz[t]\}\end{bmatrix}. 
\end{gather}
where $\bz[t] \sim \cN({\bf 0}_{2N_r}, \frac{N_0}{2}\bI_{2N_r})$ is the real-valued noise vector. 
Accordingly, we also convert the one-bit quantized signal into a real-vector form as
\begin{align}
    \by[t] & = \cQ(\br[t]) \\
    &= \cQ(\sqrt{\rho}\bH^T\bs[t] + \bz[t]),
    \label{eq:observation}
\end{align}
which is composed of $2N_r$ real-valued observations of either $-1$ or $+1$.
Throughout the paper, we consider to have $2N_r$ antennas to denote the real-valued ports for ease of notation, i.e., the $i$th antenna  in the real-value representation corresponds to $y_i[t]$.

\subsection{One-Bit ML Detection with CSI}
\label{subsec:optML}

We first introduce the conventional yet optimal one-bit ML detection in the case of perfect CSI. 
We define the index set of all possible symbol vectors as $\cK=\{1,\ldots,K\}$ and use $\bs_k$ to denote the $k$th pilot symbol vector in a real-vector form.
Let $\bP^{(\beta)}\in[0,1]^{K\times2N_r}$ with $\beta\in\{-1,+1\}$ denote the matrix of likelihood functions whose scalar entry $\bP_{k,i}^{(\beta)}$ means the probability that the $i$th antenna component receives $\beta$ when the users transmit the $k$th symbol vector $\bs_k$.
Assuming uncorrelated antennas,  the likelihood probability of the one-bit quantized signal vector $\by[t]$ for a given channel $\bH$ and transmit symbol vector $\bs_k$ is given as 
\begin{align}
	\label{eq:ML}
    \bbP(\by[t]|\bH, \bs_k) = \prod_{i=1}^{2N_r}\bP_{k,i}^{(y_i[t])}.
\end{align}
{\color{black} We remark that such an assumption for the uncorrelated antenna is valid for massive MIMO systems for sub-6GHz communications. 
For some wideband systems such as millimeter wave  communications, the assumption may not hold due to the strong line-of-sight channel. 
Accordingly, incorporating the antenna correlation structure in the learning-based one-bit ML detection problem would be a desirable future research direction.}

{\color{black} Recall that the one-bit observation becomes $y_i[t] = +1$ (or $-1$) when the $i$th element of \eqref{eq:rx signal} is positive (or negative).
Since the noise $z_i[t]$ follows a Gaussian distribution $z_i[t] \sim \mathcal{N}(0, N_0/2)$, the likelihood function for the $i$th antenna element of the quantized observation $y_i[t] \in \{-1,+1\}$ with the perfect CSI can be computed as
\begin{align}
    \label{eq:p}
   \bP_{k,i}^{(y_i[t])}\! &=\!  \bbP(y_i[t]|{\bh}_i, \bs_k\!) \\ 
   \label{eq:p2}
   &= \begin{cases}
  \Phi\left(\psi_{k,i}\right)  & y_i[t] = +1 \\
  1-\Phi\left(\psi_{k,i}\right) & y_i[t] = -1
\end{cases} ,
\end{align}
where $\Phi(x) = \int_{-\infty}^{x} \frac{1}{\sqrt{2\pi}}e^{-{\tau^2}/{2}}d\tau$ represents the cumulative distribution function of a standardized Gaussian distribution, and 
\begin{align}
    \label{eq:psi}
    \psi_{k,i}=\sqrt{\frac{\rho}{N_0/2}}\bh^T_i\bs_k
\end{align}
is the effective noiseless output of the $i$th antenna in real-value representation when transmitting the $k$th symbol vector.}
Using equation \eqref{eq:ML}, {\color{black} the one-bit ML detection rule can be obtained as \cite{choi2016near,jeon2018supervised}}
\begin{align}
    \label{eq:MLD}
    k^\star[t] = \argmax_{k \in \cK} \prod_{i=1}^{2N_r}\bP_{k,i}^{(y_i[t])}.
\end{align}
The detected real-valued symbol vector is then defined as $\hat\bs[t]$ = $\bs_{k^\star[t]}$ which can be mapped to  $\hat {\bar \bs}[t] \in \bbQ_M^{N_u}$ as detected QAM symbols by performing the reverse operation of \eqref{eq:real_x}.
Assuming an equal probability of $\frac{1}{K}$ for all possible symbol vectors,  the ML detection in \eqref{eq:MLD} is identical with the optimal maximum a posteriori probability detection.
We note that the optimal ML detection in \eqref{eq:MLD}  requires perfect CSI for computing \eqref{eq:p}.
The channel estimation, however, can be greatly burdensome in massive MIMO systems and much less accurate for receivers employing one-bit ADCs.
In this regard, it is desirable to perform the optimal detection without requiring explicit channel estimation in one-bit massive MIMO systems.
{\color{black} Note that the maximum number of users for multiuser MIMO in the uplink communications is specified as $12$ users (one layer per user case) in the 3GPP standard \cite{3gpp.21.915}. 
In addition, $4$-layer multiuser MIMO is commonly considered in practice. 
Accordingly, although reducing the search space is a desirable research direction in the implementation perspective, it is considered to be feasible to perform the current detection.}

\section{Preliminary:\\Naive One-bit ML Detection without CSI}
\label{sec:ML_learning}

Now, we outline a direct learning-based one-bit ML detection strategy that does not require channel estimation. 
Although this approach still requires $N_{\sf tr}$ training sequences, the learning principle is greatly simpler than the one-bit channel estimation, thereby providing robust detection performance.
Each pilot symbol vector $\bs_k\in\bbQ_M^{Nu}$ is transmitted $N_{\sf tr}$ times throughout the pilot transmission of length $N_t$. 
The BS  aims to approximate the true likelihood probability $\bP_{k,i}^{(\beta)}$ by observing the frequency of $y_i[t] = +1$ and $y_i[t] = -1$ {\color{black}during the transmission of the $k$th symbol vector as
\begin{align}
    \label{eq:p1_learning}
    \hat \bP^{(\beta)}_{k,i}\! =\! \begin{cases}  \hat \bP^{(+1)}_{k,i} = \frac{1}{N_{\sf tr}}\sum\limits_{\tau=1}^{N_{\sf tr}} \bb1{\{y_i[(k-1)N_{\sf tr} + \tau] = +1\}}\vspace{0.4em}\\ 
     \hat \bP^{(-1)}_{k,i} =  1- \hat \bP^{(+1)}_{k,i} 
    \end{cases}
\end{align}
where  $k \in \cK$ and $\beta \in \{+1,-1\}$ is the one-bit observation.
\begin{remark}
\normalfont
The operation in \eqref{eq:p1_learning} counts the number of $+1$'s at the $i$th antenna element out of the $N_{\sf tr}$ consecutive observations triggered by $\bs_k$.
Since each observation $y_i[t]$ follows independent and identically distributed (IID) Bernoulli variable with a probability of $\Phi\left(\psi_{k,i}\right)$, $\hat \bP^{(+1)}_{k,i}$ can be interpreted as a binomial distribution averaged by $N_{\rm tr}$, $\cB\left(N_{\sf tr},\Phi\left(\psi_{k,i}\right)\right)/N_{\sf tr}$, which approaches $\hat \bP^{(+1)}_{k,i} \to \Phi\left(\psi_{k,i}\right)$ as $N_{\sf tr}$ goes infinity. 
\end{remark}}

{\color{black}
We note that the amount of training can be cut in half by making
$\bs_k$ and $\bs_{k+\frac{K}{2}}$ symmetrical about the origin for $k\in\{1,\ldots,\frac{K}{2}\}$, i.e., $\bs_k=-\bs_{k+\frac{K}{2}}$.
By \eqref{eq:p2} and \eqref{eq:psi}, we can establish that 
\begin{align}
    \bP_{k+\frac{K}{2},i}^{(-1)} &= 1-\Phi\left(\sqrt{\frac{\rho}{N_0/2}}\bh^T_i\bs_{k+\frac{K}{2}}\right) \\
     &= 1-\Phi\left(-\sqrt{\frac{\rho}{N_0/2}}\bh^T_i\bs_k\right) \\
     & = \Phi\left(\sqrt{\frac{\rho}{N_0/2}}\bh^T_i\bs_k\right) \\
     & = \bP_{k,i}^{(+1)}.
\end{align}
Therefore, we can accommodate $\bP_{k,i}^{(\beta)}=1-\bP_{k+\frac{K}{2},i}^{(\beta)}$ for $k\in\{1,\ldots,\frac{K}{2}\}$ and $\beta\in\{-1,+1\}$, hence training for $k\in\{\frac{K}{2}+1\,\ldots,K\}$ is considered to be redundant.
}
After learning the likelihood functions, the BS  obtains the estimate of the likelihood probability for a given data signal $\by[t]$ as 
\begin{align}
    &\bbP(\by[t]|\bH,\bs_k) = \nonumber \\
    &\quad \prod_{i=1}^{2N_r}\!\!\Big(\hat \bP^{(+1)}_{k,i}\bb1\{y_i[t] \!=\! +1\} \!+ \!\hat \bP^{(-1)}_{k,i}\bb1\{y_i[t] \!=\! -1\}\!\Big), \label{eq:Py_learning} 
\end{align}
and the receiver can perform the ML detection presented in \eqref{eq:MLD} by searching the best index that maximizes \eqref{eq:Py_learning} over $\cK$, which yields the symbol vector with the highest likelihood of transmission, given the observed one-bit quantized measurements.

Although such one-bit ML approaches can provide a near-optimal detection performance with simple function learning techniques, they may suffer from critical performance degradation due to a limited amount of training {\color{black} which results in a zero-valued likelihood function (equivalently, also one-valued likelihood function), called under-trained likelihood function.

For the training of the $k$th symbol vector, the $i$th output provides $N_{\sf tr}$ different realizations of $\cQ\left(\sqrt{\rho}\bh^T_i\bs_k+z_i\right)$. 
Accordingly, in order to prevent the pair of $\{k,i\}$ from becoming under-trained, 
the sign of $\sqrt{\rho}\bh^T_i\bs_k+z_i$ where $z_i\sim\cN(0, N_0/2)$ needs to change 
at least once out of $N_{\sf tr}$ pilot transmissions.
However, at the high SNR regime, the sign change occurs with low probability, which leads to the $N_{\sf tr}$ quantized outputs at each antenna observed repeatedly to be either all $+1$'s or all $-1$'s due to the low power of the aggregate noise.
This phenomenon results in obtaining a number of zero-valued empirical likelihood functions in \eqref{eq:p1_learning}, e.g., $\hat \bP^{(\beta)}_{k,i} = 0$.
In other words,  the one-bit quantized observations at the high SNR regime become quasi-deterministic such that it is difficult to observe a change in the sign of the quantized output sequences during the $N_{\sf tr}$ transmissions of the symbol vector $\bs_k$. 
The problem of the under-trained likelihood function is stated in the following remark:
\begin{remark}[Under-trained likelihood functions]
\label{rm:under-trained}
\normalfont
Under-trained likelihood functions cause a significant degradation of the ML detection which uses \eqref{eq:p1_learning}.
Firstly, the likelihood computation in \eqref{eq:Py_learning} can be completely negated by any zero probability.
Secondly, when the SNR is very high, the quantized output becomes
\begin{align}
    \nonumber
    \by[t]  = \cQ(\br[t]) \approx \cQ(\sqrt{\rho}\bH^T\bs[t]) = {\rm sign}(\sqrt{\rho}\bH^T\bs[t]).
\end{align}
There can exist some symbols whose quantized outputs are equal without the noise, which means, ${\rm sign}(\sqrt{\rho}\bH^T\bs_k[t]) = {\rm sign}(\sqrt{\rho}\bH^T\bs_\ell[t])$, $k \neq \ell$.
In this case, if all likelihood functions are under-trained,  both the likelihood probabilities of ${\bf s}_k$ and ${\bf s}_\ell$ computed from  \eqref{eq:Py_learning} are highly likely to become $\bbP(\by[t]|\bH,\bs_k)=\bbP(\by[t]|\bH,\bs_\ell)=1$ in the very high SNR, which fails the detection for such symbols.
A similar issue also  occurs for conventional one-bit ADC systems with  true likelihood functions in the very high SNR.
For the learning-based approach, however, it happens even in the medium SNR due to the under-trained likelihood functions.
These problems are the key motivation of our work to prevent the under-trained functions with a short pilot length for any SNR regime.
\end{remark}
}

\begin{figure}[!t]
    \centering
    \includegraphics[width=1\columnwidth]{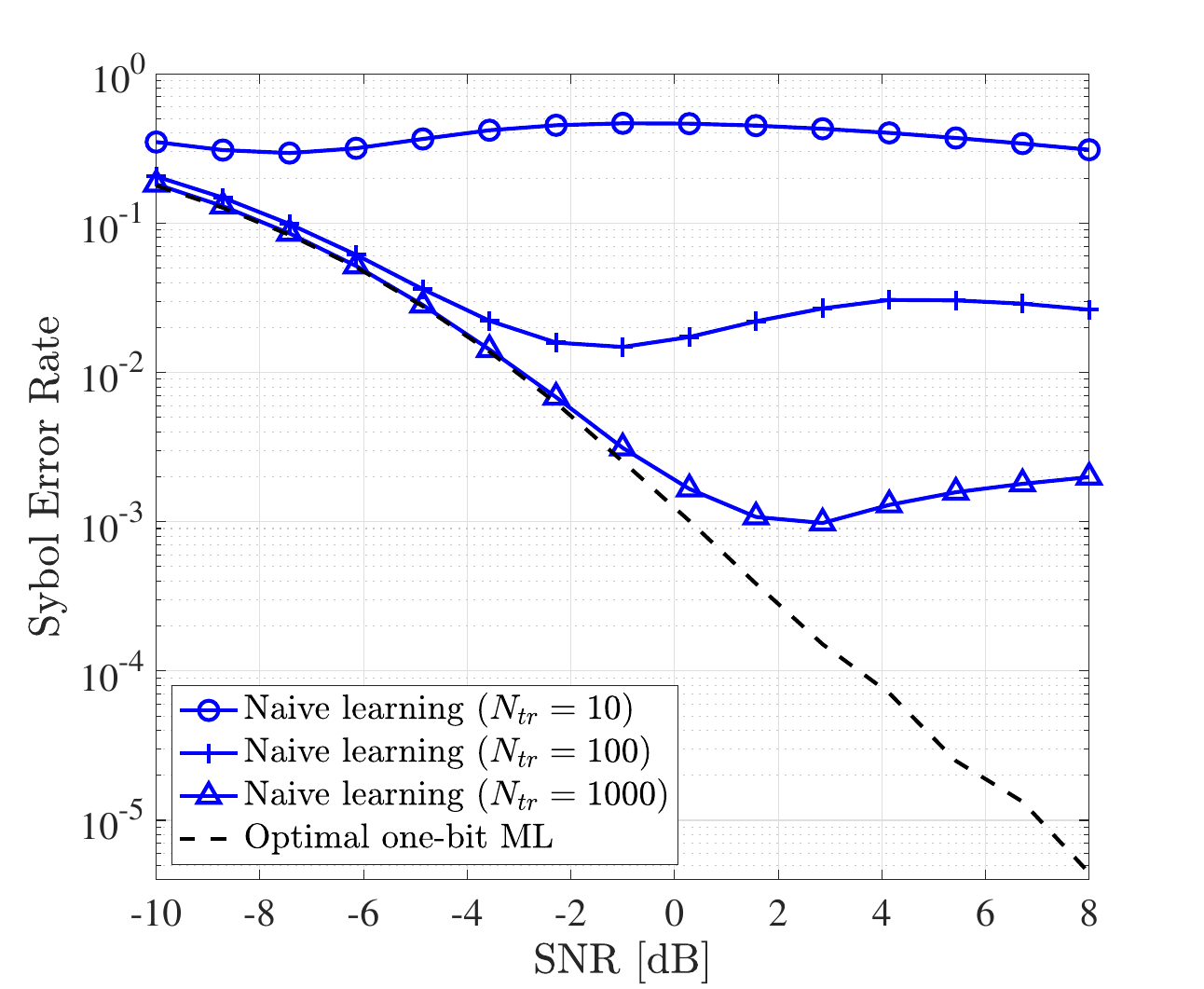}
    \caption{Symbol error rate simulation results of the optimal one-bit ML detection with full CSI against naive learning-based one-bit ML detection for $N_r =32$ receive antennas, $N_u = 3$ users, 4-QAM, and $N_{\sf tr} \in \{10, 100, 1000\}$ pilot signals.}
    \label{fig:ML_naive}
\end{figure}

Fig.~\ref{fig:ML_naive} illustrates the symbol error rates (SERs)  of the optimal one-bit ML detection and  the naive approach with the number of training samples $N_{\sf tr} \in \{10, 100, 1000\}$ for $N_r = 32$ receive antennas, $N_u =3$ users, and $4$-QAM constellation with respect to the SNR.
{\color{black}
Regarding the statistics of the channel, we consider Rayleigh fading with zero mean and unit variance.
}
It is observed that although we increase the number of pilot signals, the naive approach starts to become increasingly problematic at the medium to high SNR since the under-trained likelihood functions start to appear more frequently as the SNR increases.
Therefore, this critical limitation of the naive learning-based approach needs to be resolved to deploy the one-bit ADC systems in practice.
The main challenge lies in ensuring the robustness of learning-based ML detection to the training duration $N_{\sf tr}$ across all SNR ranges.

\section{Adaptive Statistical Learning without CSI}
\label{sec:ML}

In this section, we present an adaptive learning-based ML detection method for uplink MIMO systems with one-bit ADCs in order to closely achieve the optimal CSI-aware ML detection performance without suffering the error floor of the naive learning approach observed in Fig.~\ref{fig:ML_naive} and without the need for explicit channel estimation.
Being identical to the maximum a posteriori estimation, the ML estimation is optimal in minimizing the probability of detection error when all possible transmit symbols have an equal probability of being transmitted.
Accordingly, the proposed method can attain near-optimal detection performance without requiring explicit channel estimation.



\subsection{Dither-and-Learning}
\label{subsec:DL}

\begin{figure}[!t]
    \centering
    \includegraphics[width=1\columnwidth]{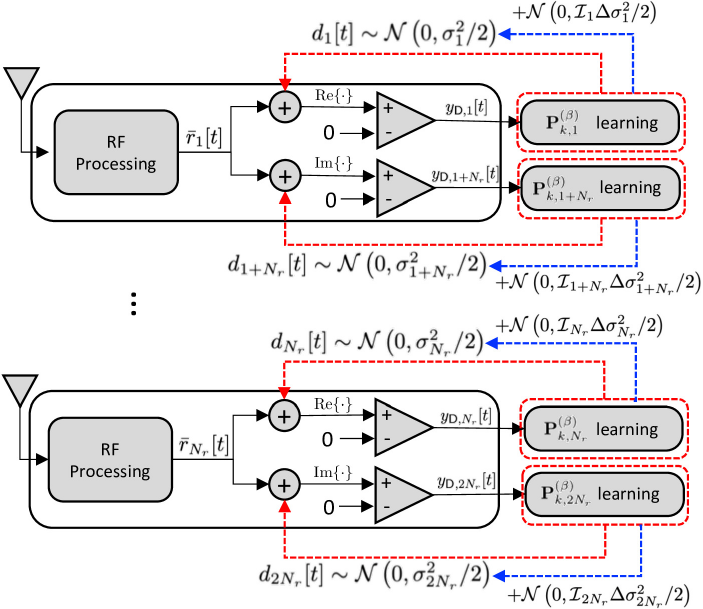}
    \caption{Illustration of the base station architecture with one-bit ADCs for $t\in\{\left(k-1\right)N_{\sf tr}+1,\ldots,kN_{\sf tr}\}$ for the training the $k$th symbol vector. {\color{black}Signals after ADCs are in real-value representation.} 
    During the pilot transmission phase, dithering signals are added before the quantization block. 
    Based on the feedback information, the statistics of the dithering signal is updated.}
    \label{fig:receiver}
\end{figure}

To resolve the problem of the under-trained likelihood functions, we propose the dither-and-learning method that can learn the likelihood functions with a reasonable training length $N_{\sf tr}$.
As shown in Fig.~\ref{fig:receiver}, the BS appends antenna-wise dithering signals $d_i[t]$ to the analog baseband received signal $r_i[t]$ during the training phase.
After dithering, the quantization input at time $t$ in the real-vector form becomes
\begin{align}
    \br_{{\sf D}}[t] &= \br[t] + \bd[t]\\
    & = \sqrt{\rho}\bH^T\bs[t] + \bz[t] + \bd[t].
\end{align}
We let $\sigma_{i}^2/2$ denote the variance of the real-valued dithering signal at the $i$th antenna and consider $\bd[t] \sim \mathcal{N}({\bf 0}_{2N_r}, \bSigma)$ where $\bSigma = {\rm {diag}}(\sigma_{1}^2/2,\dots,\sigma_{2N_r}^2/2)$.
The distribution of the dithering signal is controlled at the BS.
The dithered and quantized signal associated at time $t$ becomes 
\begin{align}
    \by_{{\sf D}}[t] & = \cQ(\br_{{\sf D}}[t]) \\
    &= \cQ(\sqrt{\rho}\bH^T\bs[t] + \bz[t] + \bd[t])\in\{+1,-1\}^{2N_r}.
\end{align}
By injecting the dithering signal $\bd[t]$ into the unquantized signal $\br[t]$, we allow the dithered signal $\br_{{\sf D}}[t]$ to cross the decision threshold within a limited amount of learning, thereby avoiding under-trained likelihood functions and facilitating the acquisition of statistical patterns.
The dithering signal is used only for the training purpose as stated in Remark~\ref{rm:ditheringnoise}.

{\color{black}
\begin{remark}[Dithering noise]
\label{rm:ditheringnoise}
\normalfont
The artificial dithering signals are added during the training phase to promote the change of sign of received signals $r_i[t]$ for a given pilot symbol within $N_{\rm tr}$ pilot transmissions.
As described in Remark~\ref{rm:under-trained}, it is important to capture a change of sign within $N_{\sf tr}$ pilot signals. 
By adding the dithering noise, we obtain $N_{\sf tr}$ different realizations of $\cQ\left(\sqrt{\rho}\bh^T_i\bs_k+z_i+d_i\right)$. 
Then $z_i+d_i\sim\cN(0, (N_0+\sigma_i^2)/2)$ needs to be less and also larger than  $-\sqrt{\rho}\bh^T_i\bs_k$ at least once out of $N_{\sf tr}$ observations to prevent $\hat \bP^{(\beta)}_{k,i}$ from being under-trained.
Such an event is less likely for the non-dithered naive approach when the noise variance $N_0$ is small.
Accordingly, by adding the dithering signal in the proposed method, 
this event is expected to occur more often compared with the non-dithered naive approach, and hence $N_{\sf tr}$ can be apparently shortened in the perspective of avoiding under-trained likelihood probabilities.
We further note that the dithered signal then is denoised to adjust the trained likelihood function to the true SNR, and the dithering noise is not present during the data transmission phase.
\end{remark}
}

As a next step, the BS  computes the estimated likelihood function for the dithered signals $\hat\bP_{{\sf D},k,i}^{(\beta)}$ as in \eqref{eq:p1_learning} for $\beta \in \{+1, -1\}$.
Without loss of generality, let us fix $\beta = +1$ for ease of explanation. 
Then, $\hat \bP_{{\sf D},k,i}^{(+1)}$ offers an estimate of the actual likelihood functions  as shown in \eqref{eq:p2} with  increased noise power:
{\color{black} 
\begin{align}
    \label{eq:p_dither}
    \hat \bP_{{\sf D},k,i}^{(+1)} &= \frac{1}{N_{\sf tr}}\sum\limits_{\tau=1}^{N_{\sf tr}} \bb1{\{y_{{\sf D},i}[(k-1)N_{\sf tr} + \tau] = +1\}}\\
    &\approx \Phi\left(\sqrt{\frac{2\rho}{N_0 + \sigma_i^2 }}\bh_i^T\bs_k\right). 
    \label{eq:p_dither_approx}
\end{align}
Since the dithering-aided counting in \eqref{eq:p_dither} approximates \eqref{eq:p_dither_approx} that includes the impact of dither signal, we plunge into the denoising stage to extract the information of desired signal and channel only without the dithering signals.}
Assuming  $N_0$ (equivalently, SNR) is known at the BS,  the BS  can find the estimate of $\psi_{k,i}$ in \eqref{eq:psi}  by leveraging \eqref{eq:p_dither}.
Such denoising is computed as
\begin{align}
    \label{eq:effective_channel}
    \hat{\psi}_{k,i} 
    &= \sqrt{1+ \frac{\sigma_i^2}{N_0}}\Phi^{-1}\left(\hat \bP_{{\sf D},k,i}^{(+1)}\right)
\end{align}

\begin{figure*}[!t]
    \centering
    \includegraphics[width=1.4\columnwidth]{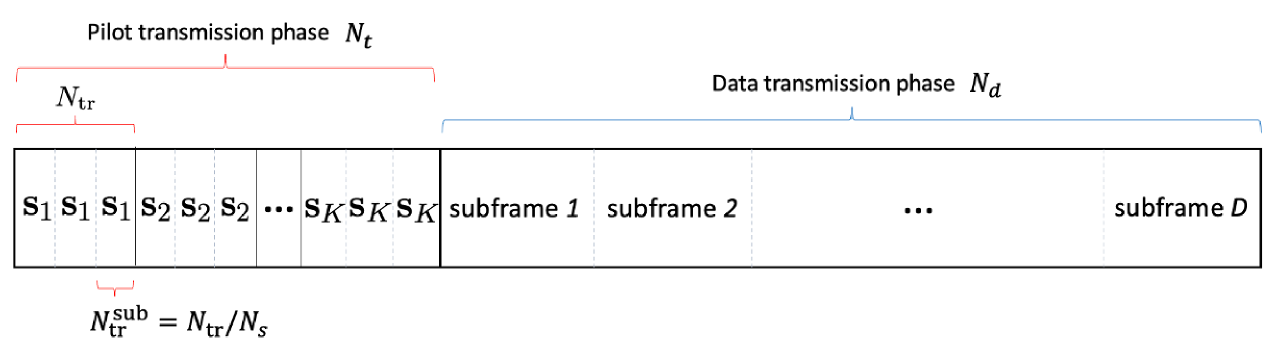}
    \caption{A communication data frame with a pilot transmission and a data transmission phases. }
    \label{fig:frame}
\end{figure*}

Finally, the BS  uses $\hat{\psi}_{k,i}$
to approximate the true (non-dithered) likelihood function $\bP_{k,i}^{(+1)}$ as 
\begin{equation}
    \tilde \bP_{k,i}^{(+1)} = \Phi\left(  \hat{\psi}_{k,i}\right).
\end{equation}
Since the likelihood function of the dithered signal $\hat \bP_{{\sf D},k,i}^{(+1)}$ in \eqref{eq:p_dither} is much less likely to have zero probability compared with that of the non-dithered case, the BS  can learn the majority of the likelihood functions $\tilde \bP^{(+1)}_{k,i}$ with a reasonable training length. 
When we observe zero likelihood functions after the dither-and-learning process, we set a very small probability that is lower than any of the non-zero likelihood functions, i.e.,
\begin{align}
     \tilde \bP_{k,i}^{(\beta)} = p_{\rm min},\quad \forall i \in \mathcal{A}^0_k(\beta),
\end{align}
where $p_{{\rm min},k} < \min_{j \in \mathcal{A}^{\rm nz}_k(\beta)}  \tilde \bP_{k,j}^{(\beta)},\ \forall \beta$,
$\mathcal{A}^0_k(\beta)$ indicates the index set of zero-valued likelihood functions for $\bs_k$ and $\beta$, and  $\mathcal{A}^{\rm nz}_k(\beta)$ is the index set of non-zero likelihood functions for $\bs_k$ and $\beta$.

For the proposed dither-and-learning method, intuitively, the power of dithering signals affects the learning performance as stated in Remark~\ref{rm:dither_power}.
\begin{remark}[Dithering power]
    \label{rm:dither_power}\normalfont
   The level of dithering power is important as insufficient dithering power continues to trigger under-trained likelihood functions, and excessive dithering power hinders recovering the symbol information, leading noise term dominant.
\end{remark}
Based on Remark~\ref{rm:dither_power}, we additionally propose an adaptive method for updating the dithering power in the subsequent section.

\subsection{Adaptive Dithering Power Update}
\label{subsec:ADL}
Using a fixed dithering variance does not suitably adjust the reception mechanism, and this behavior can cause two fundamental problems:
1) when the dithering power is low and the SNR remains high, it is highly probable to have undesirably many under-trained likelihood functions and
2) with high dithering power, although the dither-and-learning procedure successfully prevents the under-trained likelihood functions, the estimate of the effective output in \eqref{eq:effective_channel} cannot be accurate due to the large randomness of the dithering signals.
In this respect, the BS  has to properly determine dithering power considering the system environment.
To this end, we empirically update the dithering power by leveraging feedback based on the behavior of received observations and propose the adaptive dither-and-learning (ADL) method that fits the dithering power into a suitable range.

We depict the illustration of the proposed ADL method in Fig.~\ref{fig:frame}.
Rather than using up all $N_{\sf tr}$ pilot signals at once, we first divide the $N_{\sf tr}$ signals of each pilot symbol vector $\bs_k$ into $N_s$ disjoint sub-blocks in which each sub-block accommodates $N_{\sf tr}^{\sf sub}=N_{\sf tr}/N_s$ training samples where $N_{\sf tr}$ is assumed to be a multiple of $N_s$. 
Then, the $n$th dithered and quantized sub-block observed at the $i$th antenna when transmitting $\bs_k$ can be represented as
\begin{align}
    &\tilde\by_{{\sf D},k,i,n} \nonumber \\ 
    &=\Bigl\{ y_{{\sf D},i}\big[(k-1)N_{\sf tr}+\left(n-1\right)N_{\sf tr}^{\sf sub}+1\big], \nonumber \\ 
    &\;\ldots, y_{{\sf D},i}\big[(k-1)N_{\sf tr}+nN_{\sf tr}^{\sf sub}\big] \Bigr\}^T\in\{+1,-1\}^{N_{\sf tr}^{\sf sub}}, \label{eq:subblock}
\end{align}
where $n\in\{1,\ldots,N_s\}$ and $y_{{\sf D},k,i}[t]$ denotes the dithered observation at the $i$th antenna at time $t$ for the $k$th pilot symbol vector $\bs_k$. 
When the received training sequence turns out to be either $\tilde\by_{{\sf D},k,i,n} =+{\bf 1}_{N_{\rm tr}^{\rm sub}}$ or $\tilde\by_{{\sf D},k,i,n} =-{\bf 1}_{N_{\rm tr}^{\rm sub}}$ for the $i$th antenna, the dither power is regarded to be lower than the desirable dithering power for $\bs_k$ at the $i$th antenna in the currently configured system.
In such a case, we  increase the dithering noise variance of the $i$th antenna for the next sub-block by {\color{black}step size $\Delta$}\footnote{{\color{black}Here, we determine $\Delta$ through random search. To further optimize the dithering variance or $\Delta$, we may choose to apply hyperparameter tuning techniques such as Bayesian optimization \cite{snoek2012practical}. We shall leave it for a future work.}}, i.e.,
\begin{align}
    \label{eq:update}
    \sigma_{i}^2 \leftarrow \sigma_{i}^2 + \cI_i\Delta,
\end{align}
where $\cI_i$ is the indicator function defined for the $i$th antenna, i.e., $\cI_i = 1$ if $\tilde\by_{{\sf D},k,i,n} =+{\bf 1}_{N_{\sf tr}^{\sf sub}}$ or $\tilde\by_{{\sf D},k,i,n} =-{\bf 1}_{N_{\sf tr}^{\sf sub}}$, and $\cI_i = 0$ otherwise.
The indicator allows that the subsequent training sequence is more likely to observe the sign change within $N_{\sf tr}^{\sf sub}$ quantized outputs thanks to the increased perturbation.
{\color{black}
Note that the antenna-wise in-place operation in \eqref{eq:update} is performed over $N_s$ sub-blocks and $\sigma_i$ is initialized for every symbol vector.
}

Upon completing all sub-blocks, the likelihood probability of symbol vector $k$ is determined by computing the mean of the likelihood probabilities for all $N_s$ sub-blocks associated with symbol vector $\bs_k$.
Algorithm~\ref{alg:ADL} summarizes the adaptive dither-and-learning (ADL) process. 
We note that the fixed dither-and-learning method in Section~\ref{subsec:DL} is the special case of the ADL method with $N_s=1$.
We also remark that the ADL method prevents not only the under-trained likelihood functions but also the undesirably large fluctuations of the received signals since  the dithering power update is supervised by the BS  to fit into the appropriate SNR region based on the observations.

\begin{algorithm}[t]
 \caption{Adaptive Dither-and-Learning (ADL)}
 \label{alg:ADL}
{Initialize} $\tilde \bP_{k,i}^{(+1)}=0 \;\;\forall k,i$ \\
Fix the increment of the dithering variance, 
$\Delta$.\\
\For {$k = 1$ to $K$}{
   Initialize $\sigma_{i}^2 = \sigma^2$ and $\cI_i = 0 \;\;\; \forall i$.\\
  \For {$n=1$ to $N_s$}{
  \For {$i=1$ to $2N_r$}{
  Observe $\tilde\by_{{\sf D},k,i,n}$  \eqref{eq:subblock} during $N^{\sf sub}_{\sf tr}$ slots
  \\
  Compute $\hat \bP^{(\beta)}_{{\sf D},k,i}$ of $\tilde\by_{{\sf D},k,i,n}$ using \eqref{eq:p_dither}
  \\
  Compute $\hat{\psi}_{k,i}$ in \eqref{eq:effective_channel}
  \\
  $\tilde \bP_{k,i}^{(+1)} \leftarrow \tilde \bP_{k,i}^{(+1)}+\frac{1}{N_s}\Phi\left(\hat{\psi}_{k,i}\right)$\\
  \uIf {$\tilde\by_{{\sf D},k,i,n}= +{\bf 1}_{N_{\rm tr}^{\rm sub}}$ or $-{\bf 1}_{N_{\rm tr}^{\rm sub}}$}{
  $\cI_i\leftarrow 1$}
  \Else {$\cI_i\leftarrow 0$} 
 $\sigma_{i}^2 \leftarrow \sigma_{i}^2 + \cI_i\Delta$
  }}}
\Return{\ }{$\tilde \bP^{(+1)}$ and $\tilde \bP^{(-1)}=1-\tilde \bP^{(+1)}$}.
 \end{algorithm}


\subsection{SNR Estimation}
\label{subsec:SNRestimation}

\begin{figure}[!t]
    \centering
    \includegraphics[width= 1\columnwidth]{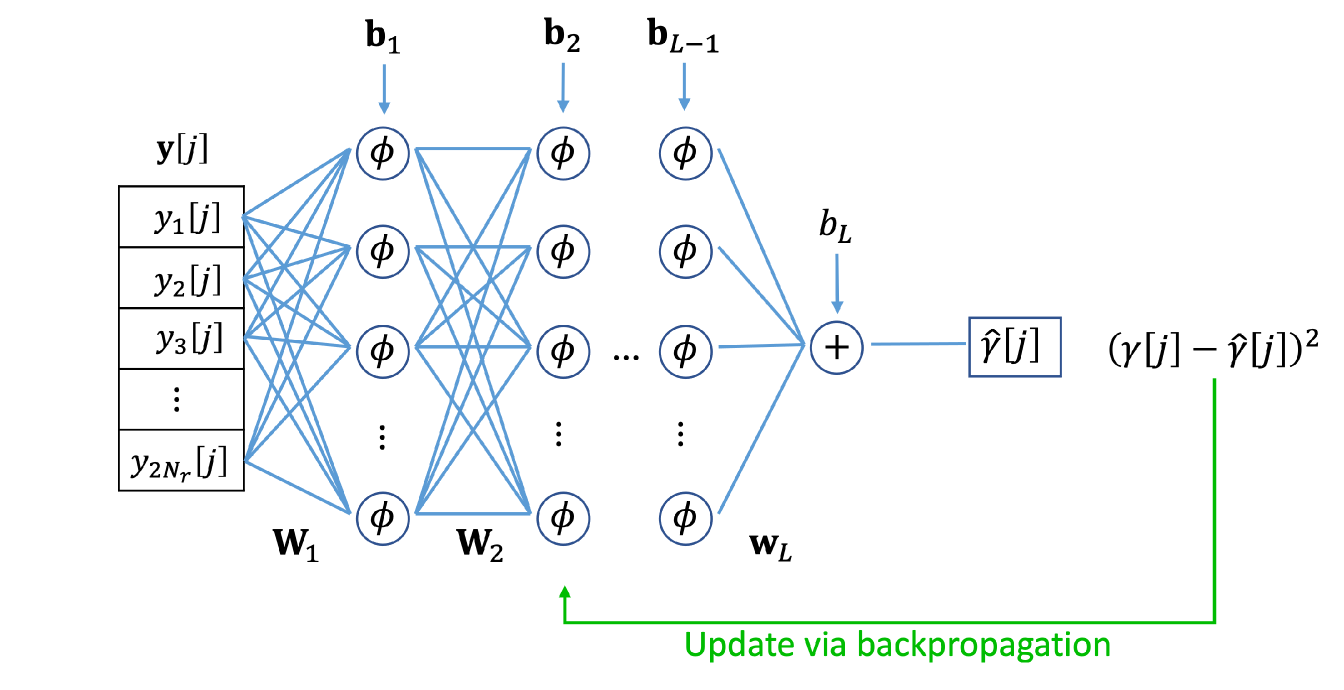}
    \caption{Illustration of the supervised offline training of the SNR using deep neural networks.
    The networks are updated in the direction of reducing estimation errors.}
    \label{fig:DNN}
\end{figure}

In spite of the properly managed dithering power, the computation of likelihood probabilities using the denoising process in \eqref{eq:effective_channel} requires knowledge of the SNR $\gamma$ or equivalently, the AWGN noise variance $N_0$. 
To address this, we also present the supervised learning approach to estimate the SNR using a DNN, as illustrated in Fig.~\ref{fig:DNN}.
During the offline training phase, we collect training data points $\{\by[j]; \gamma[j]\}$ where $\by[j]\in\{+1,-1\}^{2N_r}$ is the $j$th one-bit quantized observations and $\gamma[j]$ is the true SNR at time $j$.
Once sufficient samples are collected, the BS  selects a portion of the data points as training samples and performs the supervised  learning with $\by[j]$ as inputs and $\gamma[j]$ as outputs to be estimated.
Assuming that there exist $L$ hidden layers, the estimated SNR is represented as the scalar output of the neural network expressed as 
\begin{equation}
    \hat{\gamma}[j] = \bw_L^T \ba_{L-1}+b_L,
\end{equation}
where each intermediate vector in the DNN is defined as $\ba_{\ell} = \phi\left(\bW_{\ell}\ba_{\ell-1} + \bbb_\ell\right)$ for $\ell\in\{1,\ldots,L-1\}$ with the initial point defined as $\ba_0=\by[j]$.
Here, $\phi(\cdot)$ is an element-wise activation function such as rectified linear unit or sigmoid functions.
The DNN is updated by minimizing the estimation error, i.e., $(\gamma[j]-\hat{\gamma}[j])^2$ via backpropagation, hence estimates the SNR by extracting meaningful information of the one-bit observations such as statistical pattern and the number of $+1$'s or $-1$'s.
Throughout the paper, the DNN-based SNR estimation employs four hidden layers with output dimension of $2N_r$. 
The rectified linear unit (ReLU), i.e., $\phi(x)=\max(x,0)$ is applied to each intermediate vector. 
We note that the ReLU makes the last regression layer garner a non-negative scalar which is used for back-propagation via the Adam optimizer.

\section{Extension to Channel Coding}
\label{sec:coded}

Even though the one-bit ML detection has attractive aspects, we are still confined to the uncoded hard-decision scenarios. 
Modern communication frameworks should be paired with channel coding that exhibits an impressive gain and performance calibration; however, soft outputs are needed for the decoding perspective. 
In this section, we first introduce a frame structure to use channel coding, after that we describe how to generate soft metrics from the previously trained likelihood functions. 

\subsection{Frame Structure}
For a channel-coded communication framework, we first assume that a ($\kappa$, $\eta$) binary code with the code rate of $\kappa/ \eta$  is used throughout the paper.
At the beginning of the framework, each user $u$ then generates uncoded binary messages of length $\kappa$, denoted as $\bm_u\in\{0, 1\}^\kappa$.
By encoding the binary messages with the pre-arranged channel coding scheme, we have the codeword for $\bm_u\in\{0, 1\}^\kappa$, which is denoted as $\bc_u\in\{0, 1\}^\eta$. 
Upon generating the codeword, each user combines $q\;(=\log_2{M})$ pieces of binary information together to map the binary bits into an $M$-ary QAM symbol, and then the transmitted symbol of the $u$th user at time slot $t$ is represented as
\begin{equation}
    \bar s_u[t]=f_M\left(\{c_u[(t-1)q+1],\ldots,c_u[tq]\}\right)
\end{equation}
 where $f_M:\{0,1\}^q \rightarrow \bbQ_M$ is the constellation mapping function from $q$ binary bits to $M$-ary QAM symbols and $t\in\{1,\ldots,\eta/q\}$ where $\eta/q$ means the number of channel uses for a data subframe of each user by mapping $q$ bits into a symbol. The overall communication structure is illustrated in Fig. \ref{fig:frame}. 
 Each subframe of the data transmission phase is composed of the $N_{\sf d}^{\sf sub}=\eta/q$ channel uses, and the data transmission phase consists of the $D$ subframes, i.e., $N_{\sf d}=DN_{\sf d}^{\sf sub}$.

 
\subsection{Soft Metric}

{\color{black}In Section \ref{sec:ML}, we presented how to produce the likelihood probability utilizing the repeated transmissions with $N_{\sf tr}$ pilot signals per possible symbol vector and the ADL technique.
Furthermore, from the calculated likelihood probabilities, we can compute a likelihood ratio for a given data payload observation $\by_d[t]$. }

We note that the one-bit observation at the $t$th time slot is held accountable for the LLR computation of the $q$ positions of each user; as a result, the LLR needs to be calculated based on the user-wise and bit-wise operation.
To this end, regarding the $\ell$th  bit of the $u$th user's QAM symbol, we separate the index set of all possible symbol vectors into two non-overlapping subgroups as follows:
\begin{align}
\cS_{\ell, b}^u \triangleq \{k\; | \;\bar s_{k,u}=f_M\left(\{c_1,\ldots,c_q\}\right), c_\ell=b,  k \in \mathcal{K}\},	\label{eq:subset}
\end{align}
where $b\in\{0,1\}$ and $\bar s_{k,u}$ denotes  the $u$th element of $\bar\bs_k$ which is the QAM symbol of user $u$.
Consequently, each subset in \eqref{eq:subset} is crafted to separate $K$ indices into two disjoint sets in terms of the $\ell$th  bit of the $u$th user's bit sequence that corresponds to $\bar s_{k,u}$.
By the definition of \eqref{eq:subset}, we have $\cS_{\ell, 0}^u \cap \cS_{\ell, 1}^u=\emptyset$ and $\cS_{\ell, 0}^u \cup \cS_{\ell, 1}^u=\mathcal{K}$ for any $\ell$ and $u$.
Note that the subsets are defined regardless of current observations and computed only once when the set of system parameters is configured.

Leveraging the two separated subgroups and the pre-determined likelihood probabilities for the given observation, the corresponding LLR of the $\ell$th bit of the $u$th user at time $t$ can be represented as
\begin{align}
\nonumber
\Lambda_{(t-1)q+\ell}^u(\by_d[t]|\bH)&\stackrel{(a)}{=} \log\frac{\bbP(c_u[(t-1)q+\ell]=0|\by_d[t], \bH)}{\bbP(c_u[	(t-1)q+\ell]=1|\by_d[t], \bH)} 
\\
&\stackrel{(b)}{=}\log\frac{\bbP(\by_d[t]|c_u[(t-1)q+\ell]=0, \bH)}{\bbP(\by_d[t]|c_u[(t-1)q+\ell]=1, \bH)} \nonumber 
\\
&\stackrel{(c)}{=}\log\frac{\sum_{k\in\cS_{\ell, 0}^u}{\bbP(\by_d[t]|\bs_k, \bH)\bbP(\bs_k)}}{\sum_{k\in\cS_{\ell, 1}^u}{\bbP(\by_d[t]|\bs_k, \bH)\bbP(\bs_k)}} \nonumber
\\
&\stackrel{(d)}{=}\log\frac{\sum_{k\in\cS_{\ell, 0}^u}{\prod_{i=1}^{2N_r}\bP_{k,i}^{(y_{d,i}[t])}}}{\sum_{k\in\cS_{\ell, 1}^u}{\prod_{i=1}^{2N_r}\bP_{k,i}^{(y_{d,i}[t])}}},
\label{eq:LLR}
\end{align}
where $\ell\in\{1,\ldots,q\}$, $t\in\{1,\ldots,\eta/q\}$, $(a)$ is from the definition of LLR, $(b)$ is from Bayes' rule with the equiprobability of $\by_d$ and $\bc_u$, (c) comes from the definition of sets defined in \eqref{eq:subset}, and $(d)$ is from the equiprobability of $\bs_k$ and the ML detection rule in \eqref{eq:ML}.
Finally, the collected LLRs associated with the $u$th user, i.e., $\{\Lambda_{1}^u,\ldots,\Lambda_{\eta}^u\}$, are conveyed to a channel decoder to recover the $u$th user's message ${\hat\bm}_u\in\{0,1\}^\kappa$.
Therefore, the ADL-based estimates of the likelihood functions can be successfully used for computing the LLR of the channel decoder.

\section{Simulation Results}
\label{sec:simul}

In this section, we evaluate the performance of the proposed learning-based method in terms of the number of under-trained likelihood functions, the symbol error probability (SER) for the uncoded communication systems, and the frame error probability (FER) for the coded communication systems. 
We consider Rayleigh fading model $\bar\bH$ whose each element follows $\cC\cN(0,1)$. 
We initialize the dithering variance as $\sigma_i^2 = \rho/2$ and the increment as $\Delta = \rho/3$ for all BS  antennas in the ADL case.
\subsection{Under-trained Likelihood Functions}

\begin{figure}[!t]
    \centering
    \includegraphics[width= 1\columnwidth]{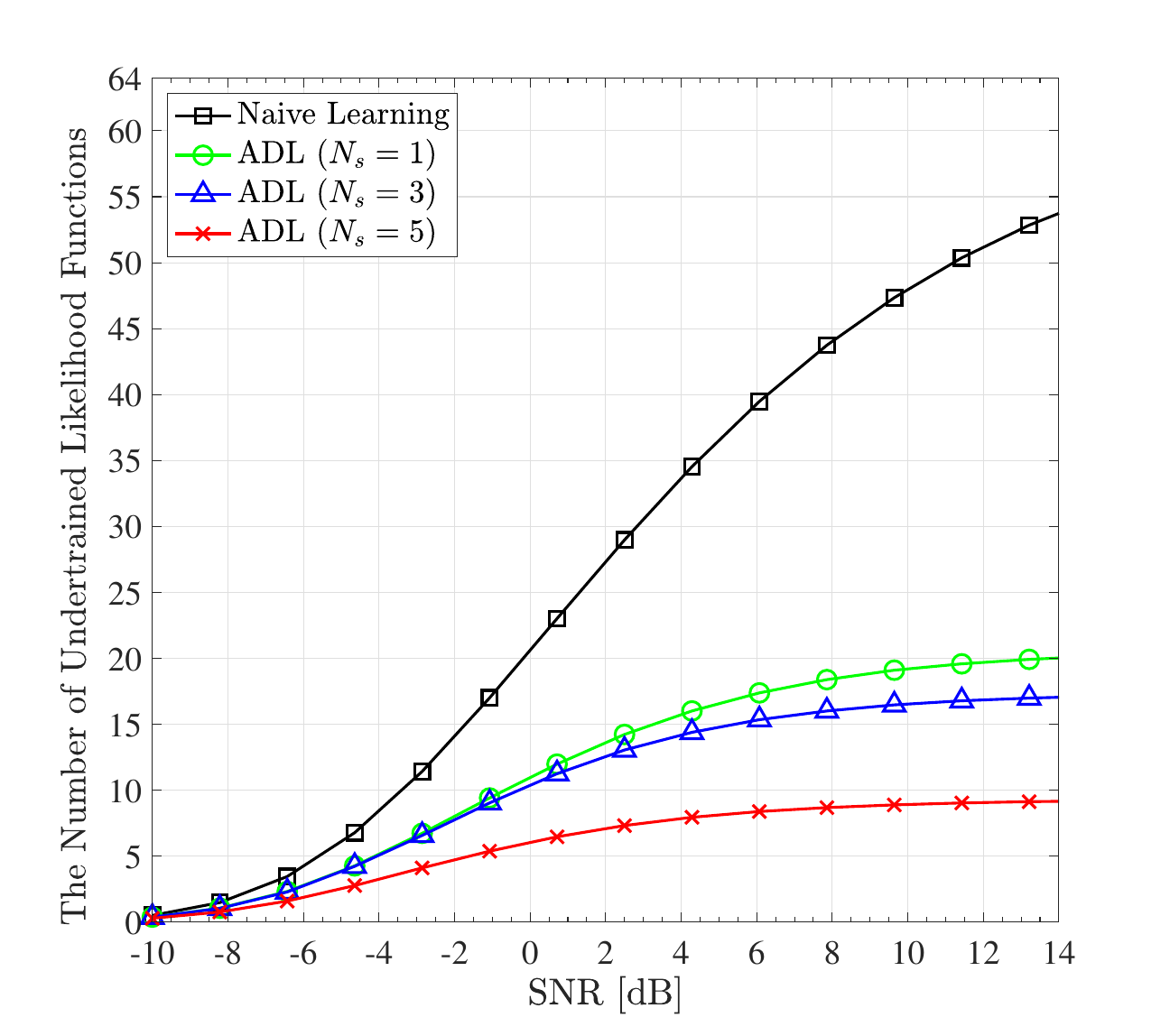}
    \caption{The number of under-trained likelihood functions among $2N_r$ likelihood functions for $N_u = 4$ users, 4-QAM, $N_r = 32$ antennas, and $N_{\sf tr}=45$ pilot signals with Rayleigh channels.
    The proposed adaptive dither-and-learning (ADL) method divides the training period into $N_s\in \{1,3,5\}$ sub-blocks for the feedback-driven update
    of dithering power.}
    \label{fig:numzero}
\end{figure}

Fig.~\ref{fig:numzero} shows the average number of under-trained likelihood functions, i.e., $\hat \bP_{k,i}^{(b)} = 0$, out of $2N_r$ antennas over the wide range of simulated SNR levels considering $N_r=32$ antennas, $N_u=4$ users, and 4-QAM.
For the learning-based detectors, we use $N_{\sf tr} = 45$ and compare the naive learning and the ADL methods with $N_s\in\{1,3,5\}$.
Recall that the ADL method with $N_s=1$ reduces to the case that uses identical and fixed dithering power without adaptation.
As the SNR increases, the number of under-trained likelihood functions for the non-dithering case rapidly approaches $2N_r$.
For the ADL case with $N_s = 1$, i.e., fixed dithering power, however, the number of under-trained likelihood functions much slowly increases with the SNR and converges to around $20$ thanks to the dithering effect.
In addition, for the ADL method with a non-trivial split factor, the number of under-trained likelihood functions increases only to $17$ and $9$ when $N_s=3$ and $N_s=5$, respectively. 
Since the ADL method decides whether to increase the dithering noise depending on the realization of each sub-block, we can further optimize the learning procedure in terms of the number of under-trained likelihood functions. 
If we properly increase $N_s$, each antenna is more likely to avoid zero-valued likelihood functions. 
As a result, with the adaptive dithering, the proposed algorithm can estimate much more valid likelihood functions, thereby increasing the detection accuracy.



\subsection{Uncoded Communication System: Symbol Error Rate}

To evaluate the data detection performance of the proposed methods in the multiuser massive MIMO system, we compare the following detection methods:
\begin{enumerate}
    \item Naive learning-based one-bit ML
    \item ADL-based one-bit ML (proposed)
    \item ADL-based one-bit ML with estimated SNR (proposed)
    \item Minimum-Center-Distance (MCD) \cite{jeon2018supervised}
    \item One-bit ZF with perfect CSI \cite{choi2015quantized}
    \item One-bit ML with perfect CSI (optimal one-bit detection)
    \item One-bit ML with estimated CSI
    \item Infinite-bit ML with perfect CSI (optimal  detection)
\end{enumerate}

\begin{figure}[!t]
    \centering
    \includegraphics[width=0.98\columnwidth]{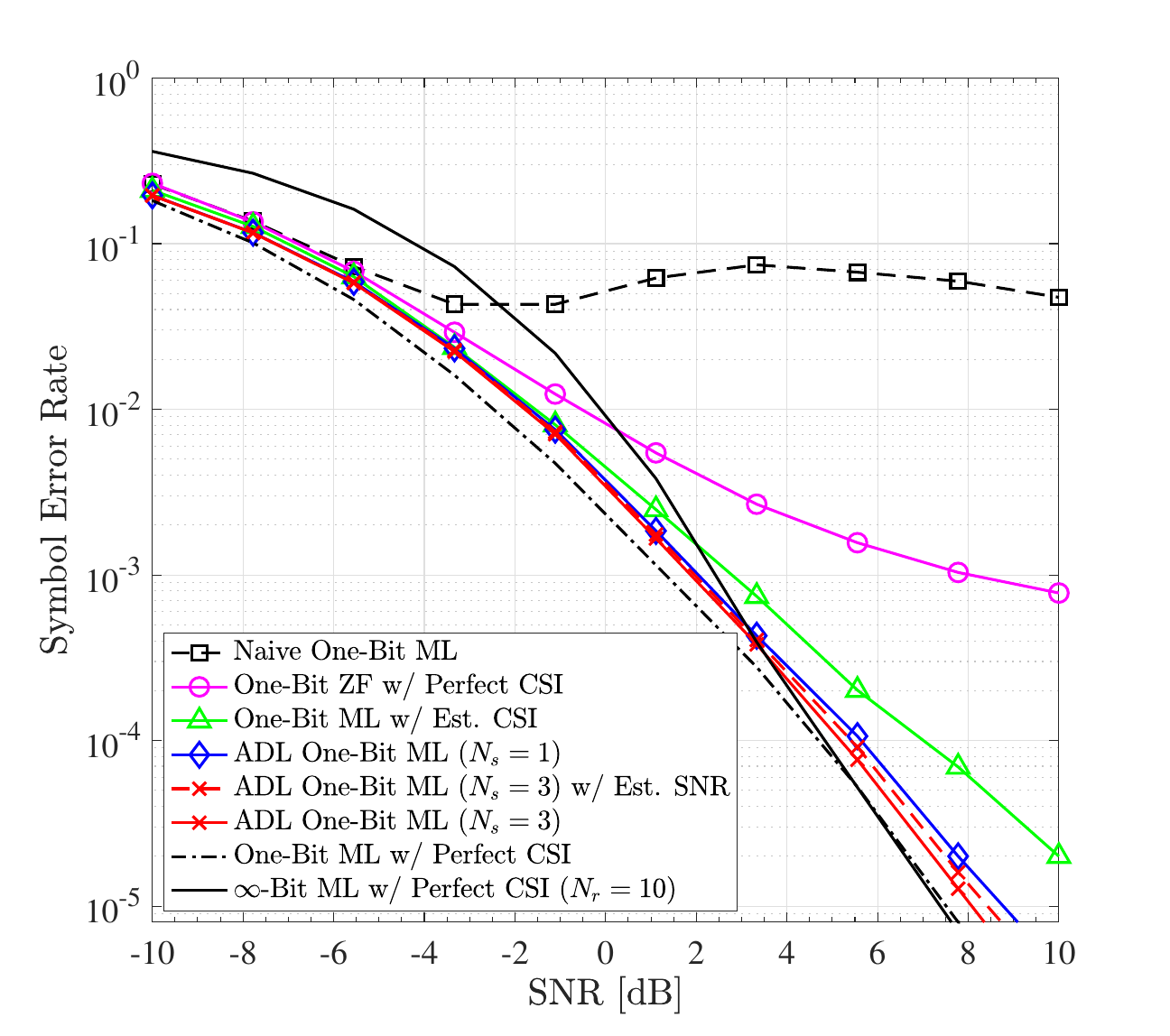}
    \caption{Symbol error rate simulation results with $N_u = 4$ users,  $N_r =32$ receive  antennas, $N_{\sf tr}=45$ training signals, and 4-QAM constellation scheme.
    The proposed adaptive dither-and-learning (ADL) uses $N_s\in\{1,3\}$ split factors.}
    \label{fig:SER45}
\end{figure}

\begin{figure}[!t]
    \centering
    \includegraphics[width=1.0\columnwidth]{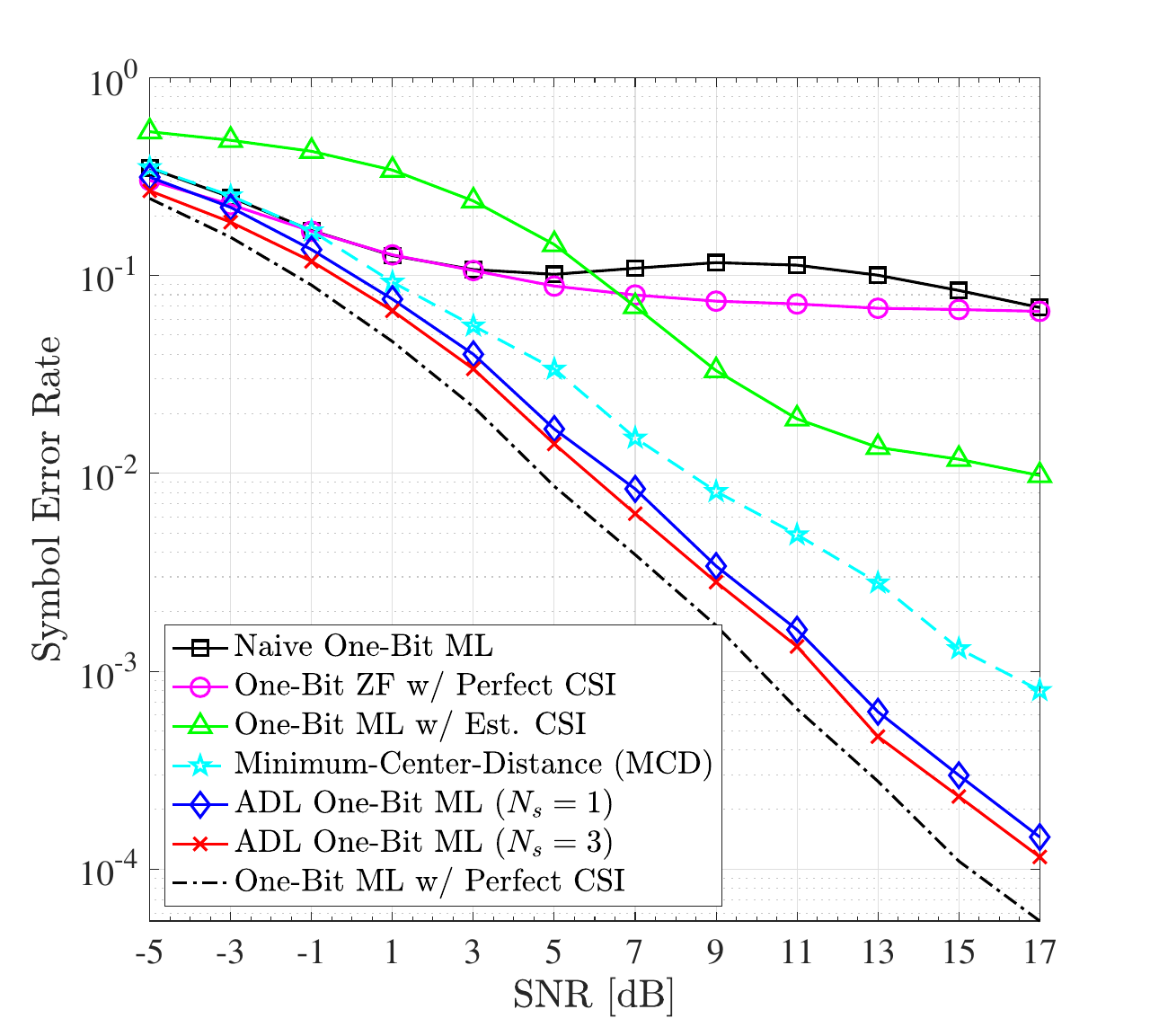}
    \caption{Symbol error rate results with $N_u = 3$ users,  $N_r =64$ BS  antennas, $N_{\sf tr}=45$ pilot signals, and 16-QAM constellation.
    The proposed adaptive dither-and-learning (ADL) method divides the training period into $N_s\in\{1,3\}$ sub-blocks.}
    \label{fig:SER16QAM}
\end{figure}

We note that the learning-based methods: 1) Naive one-bit ML, 2) ADL one-bit ML, 3) ADL one-bit ML with estimated SNR, and 4) MCD, do not require explicit channel estimation; however, the other methods either assume perfect CSI or estimated CSI at the BS.
The learning-based methods transmit $N_{\sf tr}$ pilot signals per each training symbol vector, which requires $KN_{\sf tr}$ pilot signals in total.
Accordingly, we consider that the conventional one-bit ML detection with an estimated channel also uses $KN_{\sf tr}$ pilot signals to estimate the channel.
In our simulations, the one-bit channel estimation method developed in \cite{choi2016near} is adopted to provide the estimated CSI.
For readability of the curves, we compare MCD for the 16-QAM case shown in Fig.~\ref{fig:SER16QAM}.

\begin{figure}[!t]
    \centering
    \includegraphics[width=1\columnwidth]{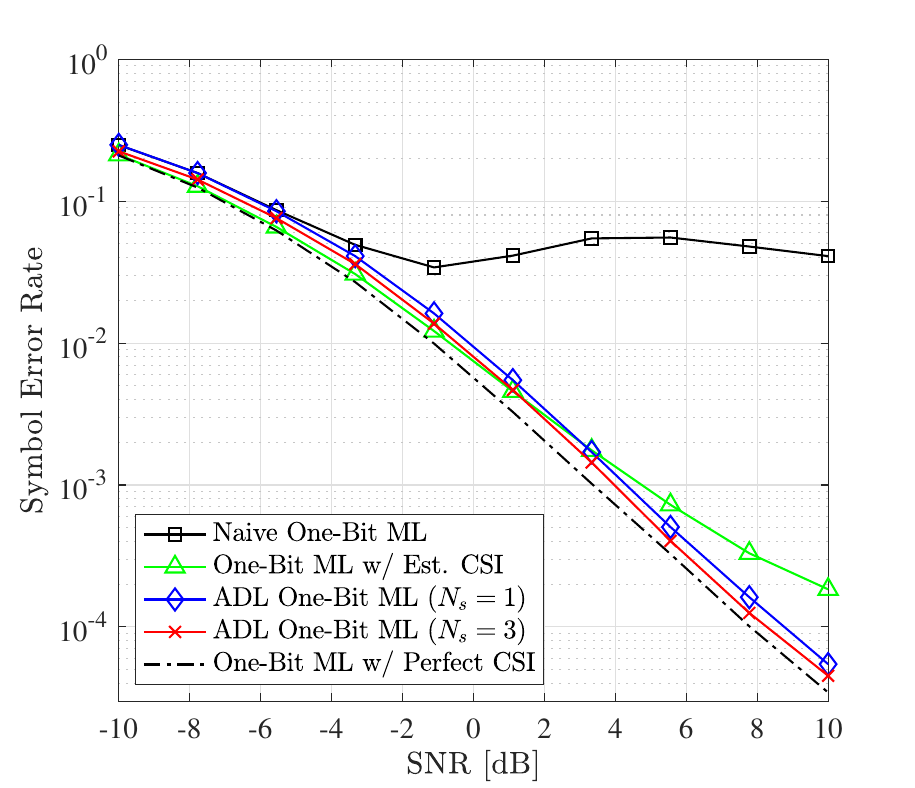}
    \caption{{\color{black}Symbol error rate simulation results with $N_u = 6$ users,  $N_r =32$ receive  antennas, $N_{\sf tr}=45$ training signals, and 4-QAM constellation scheme.
    The proposed adaptive dither-and-learning (ADL) uses $N_s\in\{1,3\}$ split factors.}}
    \label{fig:SER_UE6}
\end{figure}

Fig.~\ref{fig:SER45} presents the SER results for $N_r=32$ antennas, $N_u=4$ users, $N_{\sf tr} = 45$ pilot signals, and 4-QAM.
As expected from Fig.~\ref{fig:numzero}, the  naive-learning approach shows the catastrophic result from the medium to high SNR due to the large number of zero-valued likelihood functions.
The one-bit ZF detection which applies the pseudo-inverse matrix of the perfectly-known channel matrix onto the one-bit observations shows the large performance degradation with the error floor at the medium and high SNR regime.
The one-bit ML detection with the one-bit estimated channels  shows a larger deviation from the optimal one-bit ML detection with perfect CSI as the SNR increases due to the channel estimation error.
Unlike the above benchmarks, the proposed ADL one-bit ML methods closely follow the SER performance curve of the optimal one-bit ML case by avoiding under-trained likelihood functions as shown in Fig.~\ref{fig:numzero} and learning the likelihood functions with high accuracy. 
In addition, the proposed ADL method with $N_s=3$ has around $1.0 \ \rm {dB}$ gain over the ADL method with fixed dithering power, i.e., $N_s=1$, which demonstrates the gain of adaptive dithering based on the feedback.
We can also notice that the performance gap between the ADL method with the perfect SNR and the ADL with the estimated SNR is marginal.
This observation validates the fact that the offline supervised SNR learning can successfully capture the observation pattern to estimate the SNR required for the de-noising phase in the ADL method.
Lastly, we observe that the optimal one-bit ML detection with $N_r=32$ achieves similar target SER, e.g., $10^{-4}$ to $10^{-5}$, as the infinite-resolution ML detection with $N_r=10$ antennas.
By deploying $\sim 3\times$ more receive antennas coupled with the  low-cost one-bit ADCs, we can compensate for the severe non-linearity loss caused by one-bit ADCs and achieve higher detection performance than the infinite-bit ADC system in the low to medium SNR regime.

{\color{black}
Fig.~\ref{fig:SER_UE6} shows the SER results for $N_u=6$ users where the rest of the simulation parameters remain the same as Fig.~\ref{fig:SER45}.
We can observe that the overall SER trend of the evaluated methods is similar to the case of $N_u=4$.
The naive approach starts to suffer at the medium SNR and the channel estimation-based method that uses the same amount of training resources underperforms compared to the proposed ADL methods.
Overall, we can reaffirm that having a non-trival $N_s$ is beneficial.
}

\begin{figure}[!t]
    \centering
    \includegraphics[width=1\columnwidth]{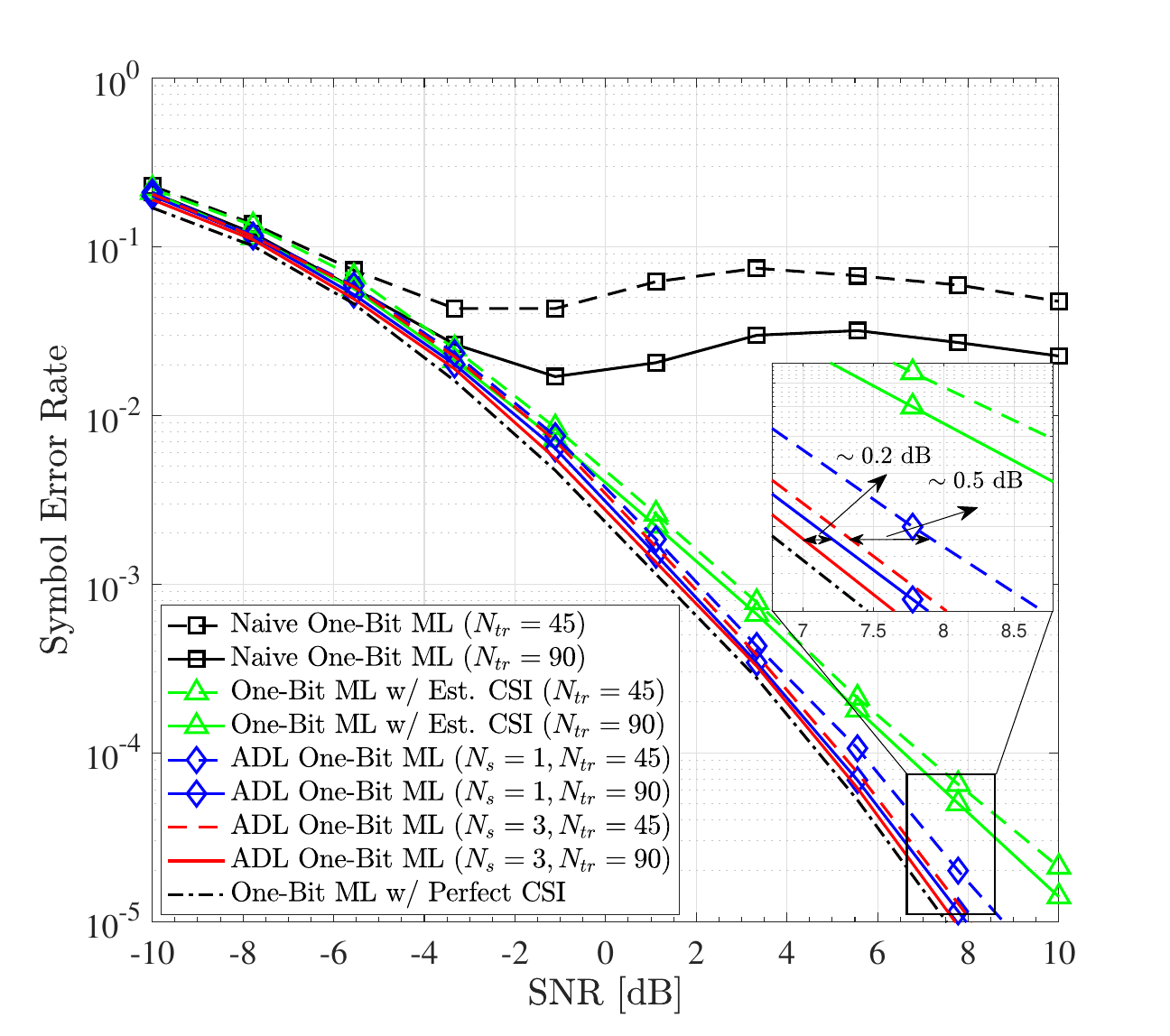}
    \caption{Symbol error rate simulation results with $N_u = 4$ users, $N_r =32$ receive antennas,  $N_{\sf tr}\in\{45,90\}$ training signals, and  4-QAM constellation.
    The proposed adaptive dither-and-learning (ADL) uses $N_s\in\{1,3\}$ split factors.}
    \label{fig:SER4590}
\end{figure}

Fig.~\ref{fig:SER4590} shows the SER performance  of the one-bit ML algorithms for different training length, $N_{\sf tr}\in\{45,90\}$ with $N_r=32$ BS  antennas, $N_u=4$ users, and 4-QAM.
We first observe that both the naive learning-based one-bit ML and the conventional one-bit ML with the estimated channel still show the noticeable performance degradation from the proposed methods for both the short and long training lengths, $N_{\sf tr}\in\{45,90\}$.
This implies that to achieve the optimal one-bit ML performance, it is necessary to use a great number of training symbols for the naive learning-based one-bit ML and the conventional one-bit ML with estimated channels.
In contrast, the proposed ADL-based one-bit ML detection offers robust performance in terms of training length.
In particular, the SER improvement of increasing $N_{s}=1$ to $N_{s}=3$ for the ADL method with $N_{\sf tr} = 90$ is about $0.2$ dB which is small compared with that for the ADL method with $N_{\sf tr} =45$. 
Therefore, we can claim that the proposed ADL method is more beneficial for the system with the limited amount of pilot signals, and using proper adaptation stages further improves the detection performance.
We can also find out that the ADL case with $N_s=3$ and $N_{\sf tr}=45$ achieves almost the same performance as the case $N_s=1$ and $N_{\sf tr}=90$, which emphasizes that adaptive learning can effectively reduce the amount of training sequences.

Fig.~\ref{fig:SER16QAM} shows the SER performance for  $N_r=64$  antennas, $N_u=3$ users, and 16-QAM. 
We use $N_{\sf tr} = 45$ training symbols for the learning-based approaches.
It is remarkable that the proposed ADL method still offers a robust detection performance whereas the one-bit ZF with perfect CSI and the one-bit ML with the estimated CSI present largely degraded detection performance.
Although the MCD method shows the lower SER than the other benchmarks, the performance gap from the proposed method is not trivial and increases with the SNR.
In this regard, the simulation results demonstrate that the proposed method outperforms the state-of-the-art one-bit detection methods, is more robust to communication environments,  
and requires shorter training sequences.





\subsection{Coded Communication System: Frame Error Rate}
We consider the MIMO configuration with $N_r=32$ antennas, $N_u=4$ users, and 4-QAM.
As a sophisticated channel coding, we adopt a rate-1/2 polar code of length 128, i.e.,  $(\kappa, \eta)=(64,128)$ and a list decoding with list size 8 is used for the decoding procedure of the polar code.
In the coded communication system, we also extend the naive learning-based one-bit ML detection to the coded system and  compare the following methods:
\begin{enumerate}
    \item Naive learning-based one-bit ML
    \item ADL-based one-bit ML (proposed)
    \item One-bit successive cancellation soft-output (OSS) \cite{cho2019one}
\end{enumerate}
For the ADL methods, we allocate a total of $N_{\sf tr} = 45$ pilot signals to each symbol vector. 
Unlike the learning-based methods, the OSS detector assumes perfect CSI to compute LLRs. 
Accordingly, it can be regarded as an FER lower bound, and we include it for providing the performance guideline.
Recall that to use state-of-the-art channel codes, we calculate LLRs using the likelihood probabilities derived by each method. 


\begin{figure}[!t]
    \centering
    \includegraphics[width=0.95\columnwidth]{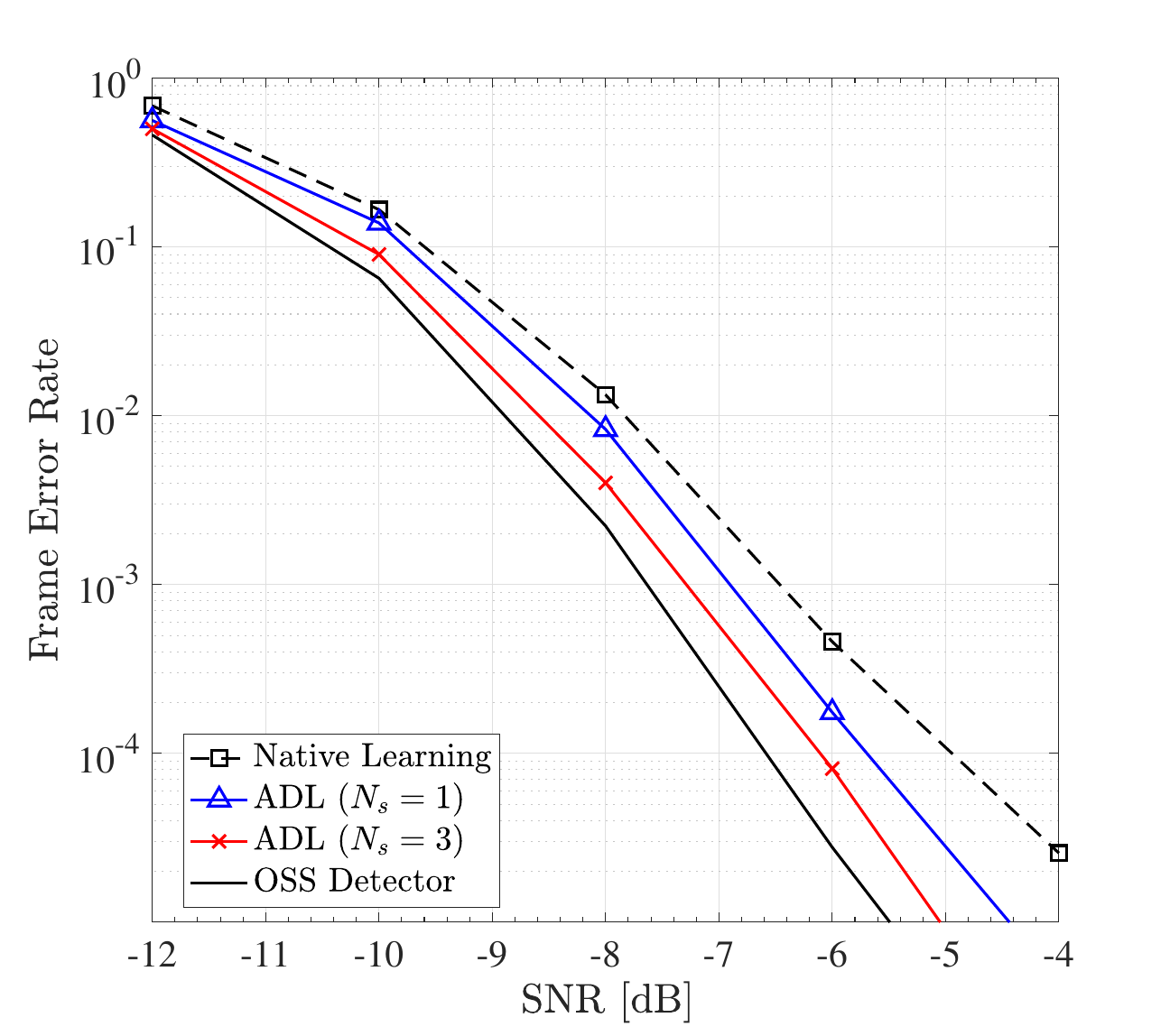}
    \caption{Frame error rate results for $N_u = 4$ users,  $N_r =32$ BS  antennas,  $N_{\sf tr}=45$, 4-QAM constellation, and a  polar code of rate $1/2$ where $(\kappa, \eta)=(64,128)$.
   The proposed adaptive dither-and-learning (ADL) method learns the likelihood probability with split factor $N_s\in\{1,3\}$.
   The one-bit successive-cancellation soft-output (OSS) detector is valid in the case of perfect CSI.}
    \label{fig:FER}
\end{figure}

Fig.~\ref{fig:FER} illustrates the FER of the channel-coded systems. 
The naive learning one-bit detection no longer experiences the tragic reverse trend shown in the uncoded systems; however, the performance gap from the proposed method grows up as SNR increases.
In addition, the FER of the ADL method with $N_s = 3$ split factor is placed between that of the OSS detector and the ADL method with $N_s=1$, thereby showing the advantage over the ADL with fixed dithering power.
Again, the ADL method with $N_s =3$ can achieve the improvement owing to the fact that the ADL method can accurately learn the likelihood probabilities by avoiding zero-valued likelihood functions even with the limited amount of training sequences. 
In summary, although the performance of the naive learning-based approach is devastated by the under-trained probabilities in the uncoded system, the likelihood probability in \eqref{eq:Py_learning} is still capable of being computed with the under-trained likelihood functions for the LLR defined in \eqref{eq:LLR} for the coded systems.
Regarding the probability learning accuracy, however, the proposed ADL method can perform better than the naive learning approach, thereby increasing the performance gap with the SNR.

\begin{figure}[!t]
    \centering
    \includegraphics[width=1\columnwidth]{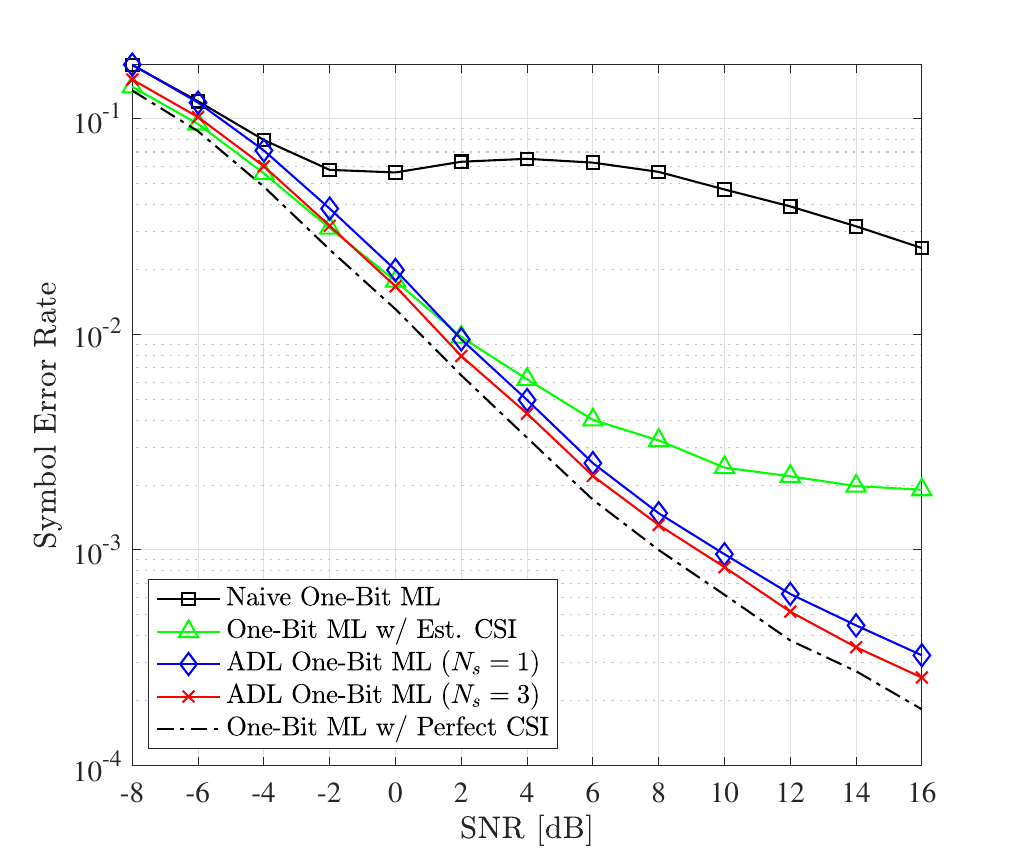}
    \caption{{\color{black}Symbol error rate simulation results with $N_u = 4$ users,  $N_r =32$ receive  antennas, $N_{\sf tr}=45$ training signals, and 4-QAM constellation scheme with geometric channels.
    The proposed adaptive dither-and-learning (ADL) uses $\sigma_i^2 = \rho/2$, $\Delta = \rho/3$, and $N_s\in\{1,3\}$ split factors.}}
    \label{fig:SER_UE4mmwave}
\end{figure}

{\color{black}
\subsection{Millimeter Wave Channel Case}
To evaluate the proposed algorithm for a mmWave channel, we adopt a geometric channel model whose number of channel paths is $L$. 
In simulations, we assume that users have the same number of paths for simplicity.
Noting that $L$ is typically small due to the limited scattering nature of  mmWave signals, the correlated mmWave channel propagated from user $u$ to the BS is expressed as \cite{li2016robust}
\begin{equation}
    {\bf \bar h}_{u} = \sqrt{\frac{N_r}{L}}\sum_{\ell=1}^{L}{\alpha_{u}(\ell)\bu(\phi_{{u}}(\ell),N_r)},
\end{equation}
where $\alpha_{u}(\ell)\sim \mathcal{CN}(0,1)$ and $\phi_{u}(\ell)\in\left[-\frac{\pi}{2},\frac{\pi}{2}\right]$ correspond to the complex path gain and the azimuth angle of arrival (AoA) of the $\ell$th path bewteen the BS and the $u$th user, respectively.
Parametrized by an azimuth angle $\phi$, $\bu(\phi, N)$ is the array response vector (ARV) for uniform linear array  which is the collection of $N$ evenly spaced phase shifts defined as
\begin{equation}
    \bu(\phi,N) = \frac{1}{\sqrt{N}}\left[1,\;e^{-j\pi\frac{2d}{\lambda}\sin{\phi}},\;\ldots,\; e^{-j\pi(N-1)\frac{2d}{\lambda}\sin{\phi}}\right],
\end{equation}
where $\lambda$ denotes the signal wavelength and $d$ is the antenna spacing.
The complex-valued channel from $N_u$ users to the BS is then $\bar\bH \in \bbC^{N_u\times N_r}$ whose $u$th row is ${\bf \bar h}_{u}^T$.

Fig.~\ref{fig:SER_UE4mmwave} shows the SER results for $N_r=32$ antennas, $N_u=4$ users, $N_{\sf tr} = 45$ pilot signals, and 4-QAM when the aforementioned geometric channel with $L=4$ is considered.
For the ADL methods, we use $\sigma_i^2 = \rho/2$ and $\Delta = \rho/3$.
We can observe that the naive approach still suffers due to the destructive under-trained likelihood functions. 
As the SNR increases, the one-bit ML detection with the one-bit estimated channels exhibits a larger deviation from the optimal one-bit ML detection with perfect CSI due to the channel estimation error.
On the other hand, the two ADL methods follow the optimal one-bit ML detector (only for the uncorrelated channel case) in the simulated SNRs and setting the non-trivial split factor of $N_s=3$ can make the ADL improve even further.
Although compared to the SER performance of the uncorrelated channels in Fig.~\ref{fig:SER45}, the overall SER is degraded due to the assumption of uncorrelated channels,  the proposed methods still achieve a meaningful SER.
}

\section{Conclusion}
\label{sec:con}

In this paper, we proposed the statistical learning-based ML detection method for uplink massive MIMO communication systems with  one-bit ADCs.
Since the performance of learning-based one-bit detection approaches can be severely degraded when the number of training samples is insufficient, the proposed method handled such challenges by injecting dithering noise to facilitate the acquisition of statistical patterns.
Without requiring explicit channel knowledge, the dither-and-learning method performed one-bit ML detection through learning likelihood functions at each antenna.
The proposed method was more robust to the number of training symbols because the adaptive randomness triggers moderate fluctuation in the change of signs of the training sequence, thereby successfully extracting the statistical pattern of one-bit quantized signals.
We further adapted dithering power to fit the BS  into the appropriate SNR region in accordance with observations.
In addition, DNN-based SNR estimation process for denoising and extension to channel-coded systems were also proposed for more practical scenarios.
Simulation results validated the detection performance of the proposed method in terms of the training amount, SER, and FER.
Therefore, the proposed method can be a potential low-power and low-complexity solution for 6G applications.

\bibliographystyle{IEEEtran}
\bibliography{Learning_1bitML}

\begin{IEEEbiography}
[{\includegraphics[width=1in,height=1.25in,clip,keepaspectratio]{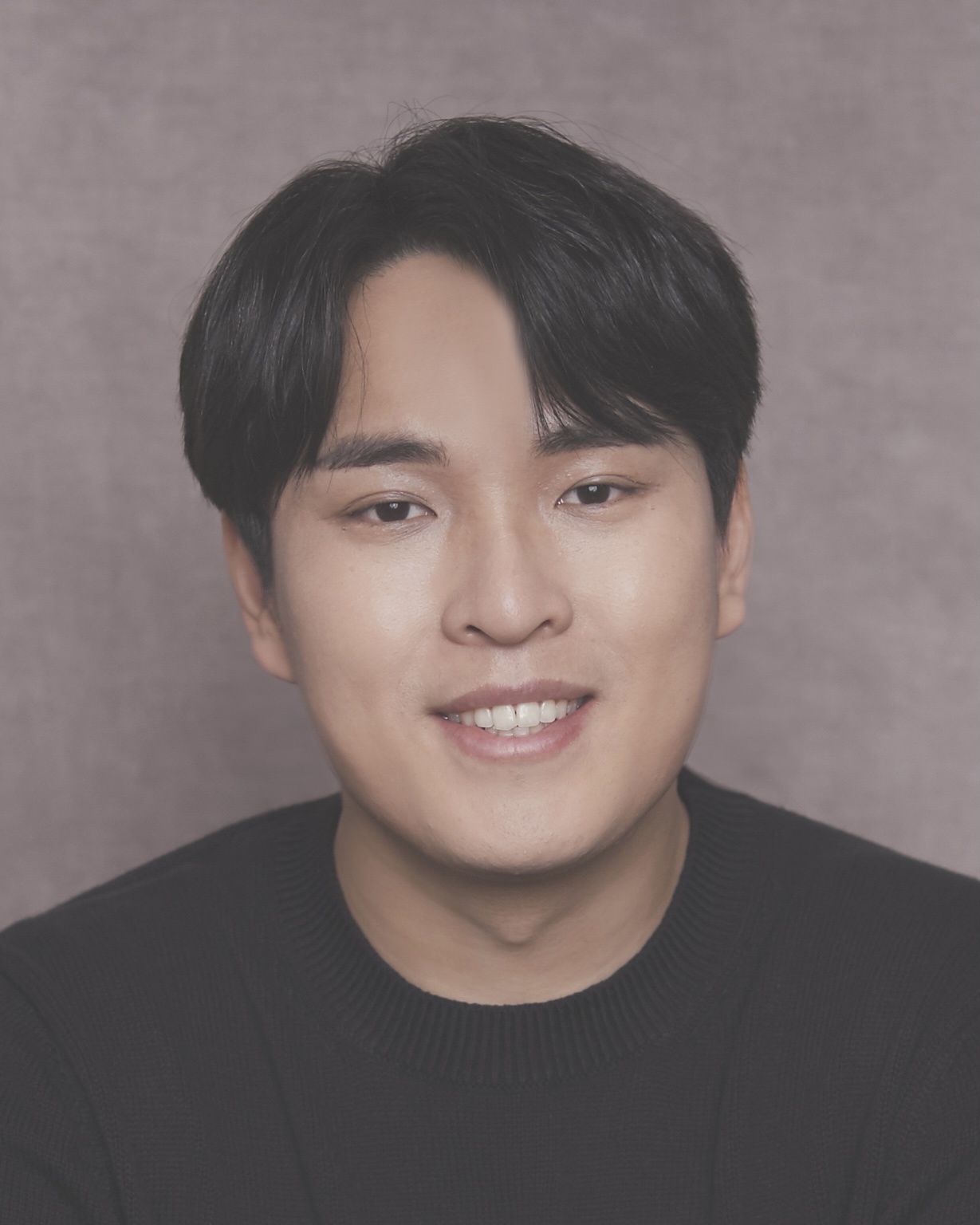}}]
{Yunseong Cho} (Member, IEEE) received his B.S. degree in electrical and computer engineering from Ajou University, South Korea in 2018. He received his M.S. and Ph.D. degrees in electrical and computer engineering from The University of Texas at Austin in 2020 and 2023, respectively. Since 2023, he has been affiliated with Samsung Research America, Plano, TX, USA, as a senior research engineer working on the developemnt of enhanced signal processing and MIMO systems to unlock Beyond 5G and 6G wireless systems. His research primarily focuses on massive MIMO,  low-resolution systems, joint transmission, user scheduling, and machine learning for communication systems.
\end{IEEEbiography}

\begin{IEEEbiography}
[{\includegraphics[width=1in,height=1.25in,clip,keepaspectratio]{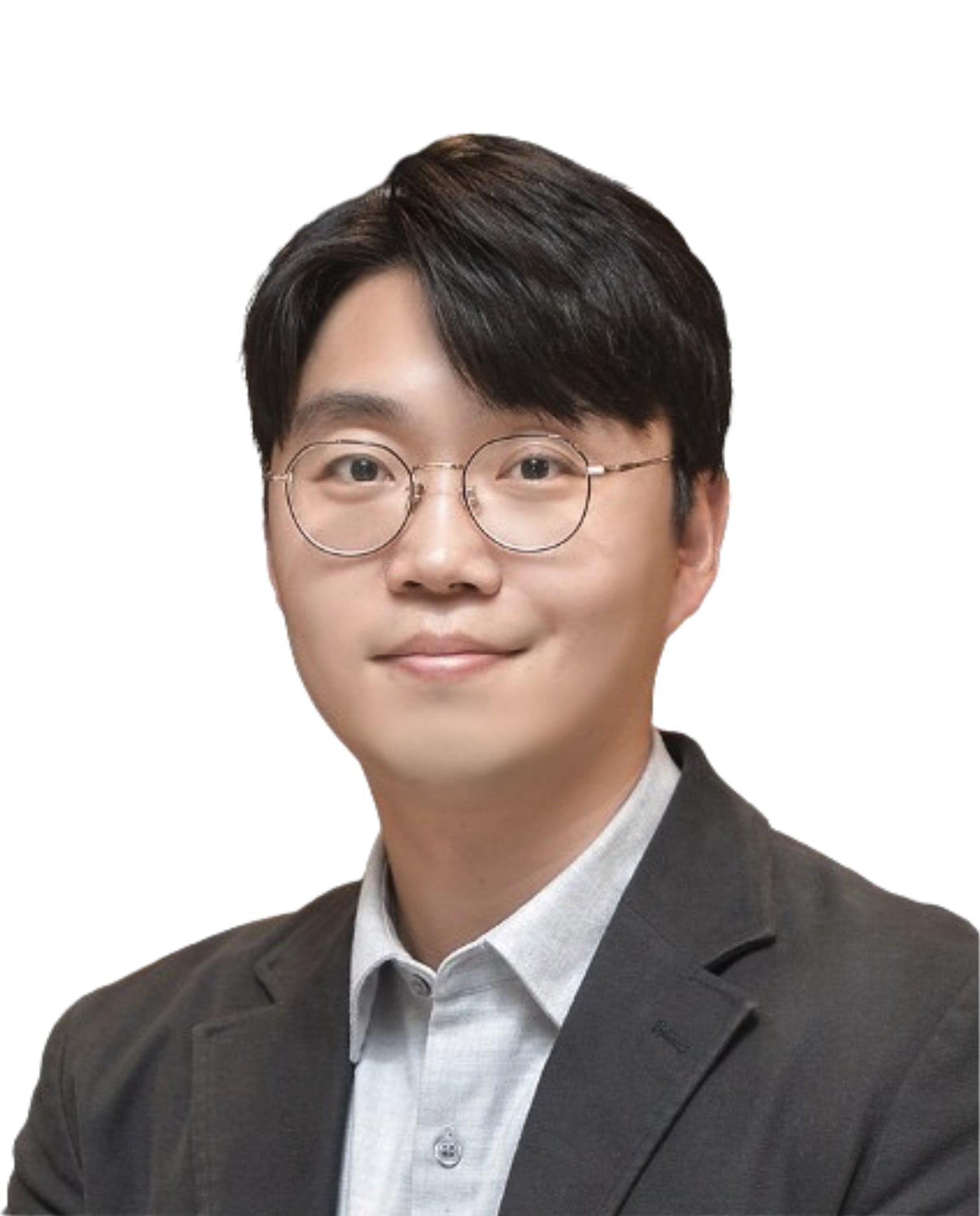}}]
{Jinseok Choi}(Member, IEEE) received his B.S. in the Department of Electrical and Electronic Engineering and B.B.A. at Yonsei University, Seoul, Korea in 2014. He received his M.S. and Ph.D. degrees in Electrical and Computer Engineering at The University of Texas at Austin, TX, USA, in 2016 and 2019, respectively. He is now an assistant professor at Korea Advanced Institute of Science and Technology (KAIST). He is also an adjunct professor of Institute for Security Convergence at KAIST, G-School at KAIST, and LG-KAIST 6G Research Center. Prior to joining KAIST, he was a senior system engineer at Wireless R\&D, Qualcomm Inc., San Diego, CA, USA, and an assistant professor at Ulsan National Institute of Science and Technology (UNIST). He received IEEE ComSoc AP Young Researcher Award in 2023 for contribution to energy-efficient communications. He serves an associate editor of IEEE Open Journal of the Communications Society. His primary research interest is to develop and analyze future wireless communication systems energy-efficient communications, new multiple access techniques, IoT communications, and satellite communications.
\end{IEEEbiography}

\begin{IEEEbiography}
[{\includegraphics[width=1in,height=1.25in,clip,keepaspectratio]{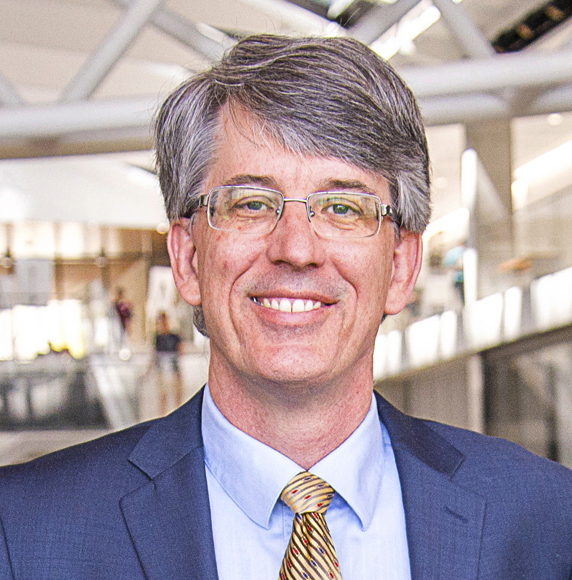}}]
{Brian L. Evans} (Fellow, IEEE) received the double major degree in electrical engineering and computer science from the Rose-Hulman Institute of Technology in 1987 and the M.S. and Ph.D. degrees in electrical engineering from the Georgia Institute of Technology in 1988 and 1993, respectively. From 1993 to 1996, he was a Post-Doctoral Researcher at the University of California at Berkeley. In 1996, he joined the faculty at The University of Texas at Austin, where he currently holds Engineering Foundation Professorship in the Department of Electrical and Computer Engineering. His research group develops algorithms with implementation constraints in mind and translates algorithms into design methods and embedded prototypes. He has published more than 280 refereed journal articles and conference papers and graduated 32 Ph.D. and 13 M.S. students. His research interests include signal processing and machine learning to increase connection speeds and reliability in wireless communication systems. He has received the three IEEE best/top conference paper awards and five teaching awards. He was the UT Austin Faculty Senate President from 2019 to 2020. He received the 2021 UT Austin Presidential Civitatis Award for “recognition of dedicated and meritorious service to the university above and beyond the regular expectations of teaching, research and service.” He received the 1997 U.S. National Science Foundation CAREER Award.

\end{IEEEbiography}

\end{document}